\pdfoutput=1

\documentclass[11pt]{article}
\usepackage[utf8]{inputenc}  
\usepackage[T1]{fontenc}      
\usepackage{CJKutf8}          
\usepackage{multirow}
\usepackage{bbding}
\usepackage{tikz}
\usepackage{booktabs}
\usepackage{pifont}
\usepackage{svg}
\usepackage[normalem]{ulem}
\useunder{\uline}{\ul}{}
\usepackage{subcaption}

\usepackage{pgfmath}

\usepackage[final]{acl}

\usepackage{times}
\usepackage{latexsym}

\usepackage[T1]{fontenc}

\usepackage[utf8]{inputenc}

\usepackage{microtype}

\usepackage{inconsolata}

\usepackage{graphicx}

\title{Flow2Code: Evaluating Large Language Models for Flowchart-based Code Generation Capability}

\author{Mengliang He\textsuperscript{1}, Jiayi Zeng\textsuperscript{1}, Yankai Jiang\textsuperscript{2}, Wei Zhang\textsuperscript{1}, \\
        {\bf  Zeming Liu\textsuperscript{3}\footnotemark[1], Xiaoming Shi\textsuperscript{1}\thanks{\ Corresponding authors: Zeming Liu, Xiaoming Shi.}, Aimin Zhou\textsuperscript{1}} \\
        \textsuperscript{1} East China Normal University, Shanghai, China \textsuperscript{2} Shanghai AI Lab, Shanghai, China \\
        \textsuperscript{3} Beihang University, Beijing, China \\
        {\tt \{51255901020, 51265901055\}@stu.ecnu.edu.cn; zhangwei.thu2011@gmail.com;} \\
        {\tt zmliu@buaa.edu.cn; \{xmshi, amzhou\}@cs.ecnu.edu.cn}
 }
 
\begin{document}
\begin{CJK*}{UTF8}{gbsn}
\maketitle
\begin{abstract}






While large language models (LLMs) show promise in code generation, existing benchmarks neglect the flowchart-based code generation. 
To promote further research on flowchart-based code generation,
this work presents Flow2Code, a novel benchmark for flowchart-based code generation evaluation. 
The evaluation dataset spans 15 programming languages and includes 5,622 code segments paired with 16,866 flowcharts of three types: code, UML, and pseudocode. 
Extensive experiments with 13 multimodal LLMs reveal that current LLMs can not generate code based on flowcharts perfectly.
Besides, experiment results show that the supervised fine-tuning technique contributes greatly to the models' performance. 
We publicly release our code and datasets at https://github.com/hml-github/Flow2Code.


\end{abstract}

\section{Introduction}
The code generation task aims to convert specific requirements into executable code~\cite{nuseibeh2000requirements}, 
which attracts interest and focus from the academic and industrial communities.
Recently, for automatic code generation, \textbf{l}arge \textbf{l}anguage \textbf{m}odels (LLMs)~\citep{openaiGPT4TechnicalReport2023a,openaiGPT4oSystemCard2024,nijkampCodeGen2LessonsTraining2023,deepseek-aiDeepSeekCoderV2BreakingBarrier2024,huiQwen25CoderTechnicalReport2024} exhibit substantial potential and show alluring application value for enhancing productivity, minimizing human error.




To comprehensively evaluate and understand the code generation capabilities of emerging LLMs, substantial efforts have been devoted to establishing and refining code generation benchmarks. 
Specifically, as shown in Figure~\ref{table:data_sum}, 
current code generation benchmarks can be classified into two categories: those based on textual descriptions, such as HumanEval~\citep{chenEvaluatingLargeLanguage2021} and MBPP~\citep{austinProgramSynthesisLarge2021}, and those based on images of programming problems or matplotlib plots, such as MMCode~\citep{li2024mmcode} and Plot2Code~\citep{wuPlot2CodeComprehensiveBenchmark2024}. 
%
Despite the potential value, these works fundamentally suffer from a critical flaw:
The lack of flowchart-based code generation evaluation.
Compared to textual descriptions, images of programming problems, or matplotlib plots, flowcharts offer a more effective and intuitive way to understand and visualize program logic such as decisions, loops, and conditionals, making them accessible to both programmers and non-programmers alike~\citep{xinogalosUsingFlowchartbasedProgramming2013}. 
Flowcharts mainly consist of three types: basic code flowcharts, Unified Modeling Language (UML) flowcharts, and pseudocode flowcharts~\citep{10.5555/1074100.1074406}.
Specifically, basic code flowcharts represent the step-by-step execution of a program, 
UML flowcharts are a formal graphical representation of an object-oriented system's structure, 
and pseudocode flowcharts are a high-level abstraction of program logic and are represented by natural Language.
Despite these advantages, there is a notable gap in code generation benchmarks, with no dedicated datasets or frameworks for generating code from flowcharts. 
This limitation hinders the ability of LLMs to fully utilize flowcharts for code generation.

\begin{table*}[t]
\resizebox{\textwidth}{!}{
\begin{tabular}{lcccccr}
\hline
Datasets & Code Flowchart & UML Flowchart & Pseudocode Flowchart & Multimodal & Multilingual & Samples \\ \hline
APPS~\citep{hendrycks2measuring} & \textcolor{red}{\ding{55}} & \textcolor{red}{\ding{55}} & \textcolor{red}{\ding{55}} & \textcolor{red}{\ding{55}}(text) & \textcolor{red}{\ding{55}} & 10,000 \\
HumanEval~\citep{chenEvaluatingLargeLanguage2021} & \textcolor{red}{\ding{55}} & \textcolor{red}{\ding{55}} & \textcolor{red}{\ding{55}} & \textcolor{red}{\ding{55}}(text) & \textcolor{red}{\ding{55}} & 164 \\
MBPP~\citep{austinProgramSynthesisLarge2021} & \textcolor{red}{\ding{55}} & \textcolor{red}{\ding{55}} & \textcolor{red}{\ding{55}} & \textcolor{red}{\ding{55}}(text) & \textcolor{red}{\ding{55}} & 974 \\
DS-1000~\citep{lai2023ds} & \textcolor{red}{\ding{55}} & \textcolor{red}{\ding{55}} & \textcolor{red}{\ding{55}} & \textcolor{red}{\ding{55}}(text) & \textcolor{red}{\ding{55}} & 1,000 \\
CodeContests~\citep{liCompetitionLevelCodeGeneration2022} & \textcolor{red}{\ding{55}} & \textcolor{red}{\ding{55}} & \textcolor{red}{\ding{55}} & \textcolor{red}{\ding{55}}(text) & \textcolor{green}{\ding{51}}(3) & 13,610 \\
MBXP~\citep{athiwaratkun2023multi} & \textcolor{red}{\ding{55}} & \textcolor{red}{\ding{55}} & \textcolor{red}{\ding{55}} & \textcolor{red}{\ding{55}}(text) & \textcolor{green}{\ding{51}}(13) & 12,425 \\
ClassEval~\citep{duClassEvalManuallyCraftedBenchmark2023} & \textcolor{red}{\ding{55}} & \textcolor{red}{\ding{55}} & \textcolor{red}{\ding{55}} & \textcolor{red}{\ding{55}}(text) & \textcolor{red}{\ding{55}} & 100 \\
CoderEval~\citep{yuCoderEvalBenchmarkPragmatic2024} & \textcolor{red}{\ding{55}} & \textcolor{red}{\ding{55}} & \textcolor{red}{\ding{55}} & \textcolor{red}{\ding{55}}(text) & \textcolor{green}{\ding{51}}(2) & 460 \\
HumanEval-X~\citep{zheng2023codegeex} & \textcolor{red}{\ding{55}} & \textcolor{red}{\ding{55}} & \textcolor{red}{\ding{55}} & \textcolor{red}{\ding{55}}(text) & \textcolor{green}{\ding{51}}(5) & 820 \\
MCEVAL~\citep{chaiMcEvalMassivelyMultilingual2024} & \textcolor{red}{\ding{55}} & \textcolor{red}{\ding{55}} & \textcolor{red}{\ding{55}} & \textcolor{red}{\ding{55}}(text) & \textcolor{green}{\ding{51}}(40) & 16,031 \\ 
BigCodeBench~\citep{zhuo2024bigcodebench} & \textcolor{red}{\ding{55}} & \textcolor{red}{\ding{55}} & \textcolor{red}{\ding{55}} & \textcolor{red}{\ding{55}}(text) & \textcolor{red}{\ding{55}} & 1,140 \\ \hline
Plot2Code~\citep{wuPlot2CodeComprehensiveBenchmark2024} & \textcolor{red}{\ding{55}} & \textcolor{red}{\ding{55}} & \textcolor{red}{\ding{55}} & \textcolor{green}{\ding{51}}(image, text) & \textcolor{red}{\ding{55}} & 132 \\
MMcode~\citep{li2024mmcode} & \textcolor{red}{\ding{55}} & \textcolor{red}{\ding{55}} & \textcolor{red}{\ding{55}} & \textcolor{green}{\ding{51}}(image, text) & \textcolor{red}{\ding{55}} & 3,548 \\ 
HumanEval-V~\citep{zhang2024humaneval} & \textcolor{red}{\ding{55}} & \textcolor{red}{\ding{55}} & \textcolor{red}{\ding{55}} & \textcolor{green}{\ding{51}}(image, text) & \textcolor{red}{\ding{55}} & 253 \\ \hline
Flow2Code & \textcolor{green}{\ding{51}} & \textcolor{green}{\ding{51}} & \textcolor{green}{\ding{51}} & \textcolor{green}{\ding{51}}(image, text) & \textcolor{green}{\ding{51}}(15) & 16,866 \\ \hline
\end{tabular}%
}
\caption{Comparison of the Flow2Code dataset with other code generation benchmarks, where the number in the Multilingual column represents the amount of programming code contained in the dataset.}
\label{table:data_sum}
\end{table*}
To address this gap, this work first introduces Flow2Code, a comprehensive code generation benchmark that includes three types of flowcharts and corresponding code in 15 programming languages.
The construction of Flow2Code consists of three parts: the creation of code and UML flowcharts, pseudocode conversion, and data checking.
After the construction of the dataset, a two-step human evaluation process is employed to ensure data quality, which includes both code verification and validation of pseudocode flowchart transformations.
Finally, the dataset is obtained with a total of 5,622 code segments, 16,866 flowcharts and includes 15 programming languages, offering a rich resource for evaluating code generation tasks.


\begin{figure*}[t]
  \includegraphics[width=1.0\linewidth]{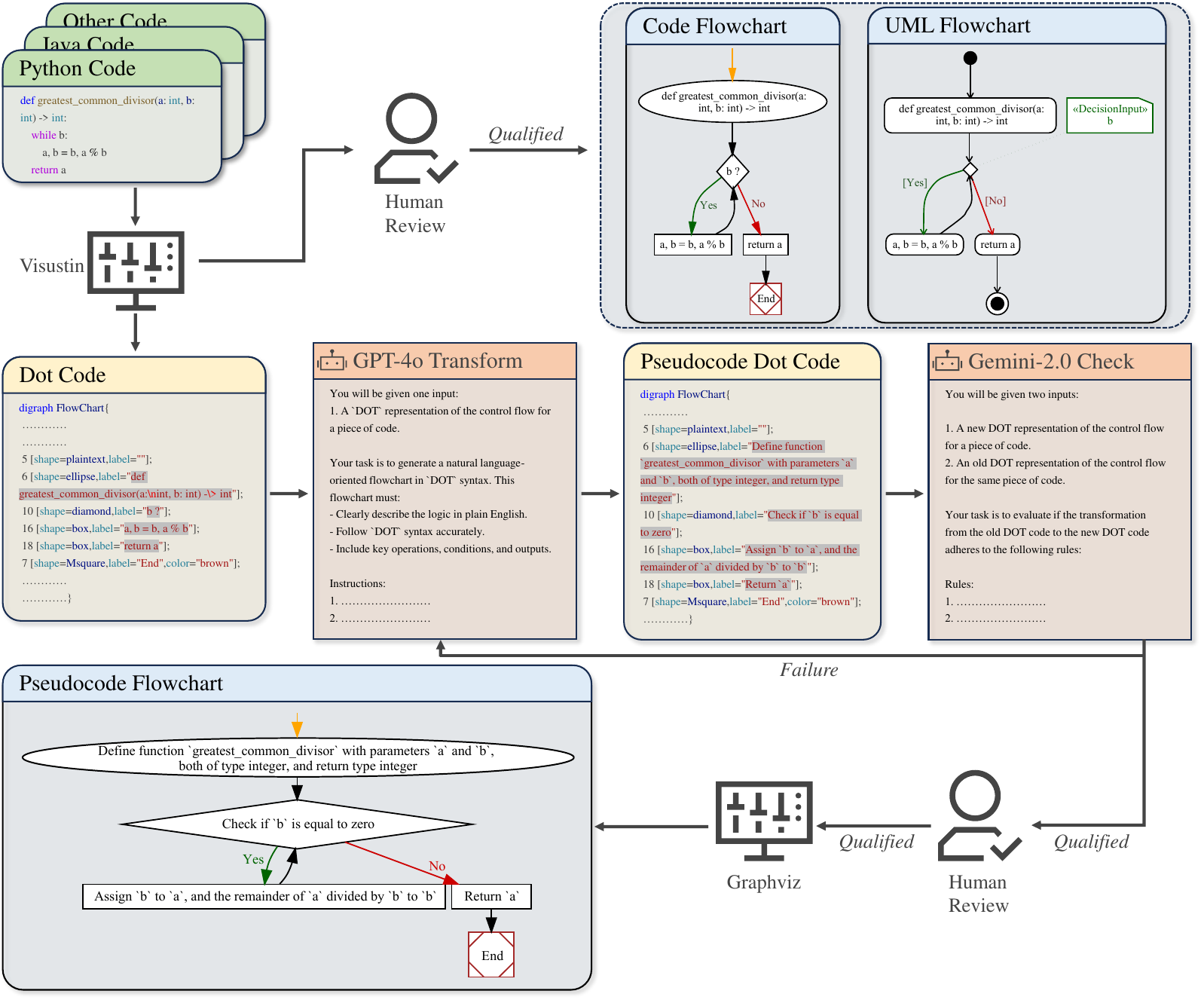}
  \caption {
  Overview of the flowchart generation process. The source code is initially converted into DOT code and code, UML flowcharts using Visustin. GPT-4o is then employed to transform the DOT code into a natural language pseudocode (highlighted in gray text). The generated DOT code is first validated through a Gemini-2.0 check, after which the pseudocode flowcharts are created. Finally, the pseudocode, code, and UML flowcharts are subjected to a comprehensive human review to ensure the accuracy and quality of the transformations.}
  \label{figure:gen-flow}
\end{figure*}
To further conduct the evaluation of current LLMs, we conduct comprehensive benchmarking experiments on the Flow2Code dataset using 13 LLMs under the settings of zero-shot and supervised fine-tuning.
The experimental results demonstrate that： 
(1) Current LLMs lack sufficient flowchart-based code generation capability, particularly on the UML and pseudocode flowcharts;
(2) Supervised fine-tuning technique is effective in improving LLMs' flowchart-based code generation.
These findings highlight areas for improvement and provide guidance for future research in code generation.

This work makes the following contributions:
\begin{itemize}
    \item We identify a key challenge for current LLMs as flowchart-based code generation.
    \item To promote further research on flowchart-based code generation, we introduce a novel benchmark, termed Flow2Code, and perform extensive evaluation experiments on various LLMs, providing a standardized platform for assessing flowchart-based code generation.
    \item Experimental results show that supervised fine-tuning on flowchart-based code generation datasets is an effective technique to improve LLMs' performance.
\end{itemize}

\section{Related work}
This Section first reviews the current research on code generation benchmarks (Section~\ref{RW:1}), followed by an overview of research on 
multimodal benchmarks (Section~\ref{RW:3}).

\subsection{Code Generation Benchmarks}\label{RW:1}
Traditional code generation benchmarks, such as HumanEval~\citep{chenEvaluatingLargeLanguage2021}, MBPP~\citep{austinProgramSynthesisLarge2021}, DS-1000~\citep{lai2023ds}, and APPS~\citep{hendrycks2measuring}, focus on text-based tasks and a single programming language. Although newer benchmarks like MBXP~\citep{athiwaratkun2023multi}, MCEVAL~\citep{chaiMcEvalMassivelyMultilingual2024}, MultiPL-E~\citep{cassanoMultiPLEScalablePolyglot2023}, humaneval-X~\citep{zheng2023codegeex}, and HumanEval-XL~\citep{peng2024humaneval} cover multiple languages and tasks, they still focus on text-based code generation. In contrast, Flow2Code incorporates flowchart-based representations, offering a more comprehensive evaluation of LLMs in multimodal code generation tasks.

\subsection{Multimodal Benchmarks}\label{RW:3}
Recent multimodal benchmarks such as MMBench~\citep{xuMMBenchBenchmarkingEndtoEnd2023}, MMMU~\citep{yueMMMUMassiveMultiDiscipline2024}, MMStar~\citep{chen2024we}, and Web2Code~\citep{yun2024web2code} evaluate MLLMs on tasks involving text and images, mainly focusing on visual reasoning and general multimodal capabilities. MMCode~\citep{li2024mmcode} and Plot2Code~\citep{wuPlot2CodeComprehensiveBenchmark2024} extend this by targeting code generation, with MMCode focusing on Python problems with visual aids and Plot2Code on plots-to-code generation. However, these datasets are domain-specific, with limited coverage of broader code generation tasks. In contrast, the Flow2Code benchmark specifically evaluates the task of translating various flowchart types—code, UML, and pseudocode—into executable code, offering a more specialized and structured evaluation across diverse programming languages.


\section{Dataset Construction}
This section first describes the data source and selection for the Flow2Code dataset (Section~\ref{DC:1}), followed by the flowchart construction process (Section~\ref{DC:2}), and concludes with an analysis of the dataset (Section~\ref{DC:3}).

\subsection{Data Selection}\label{DC:1}
The Flow2Code dataset is constructed from four key datasets: HumanEval-X~\citep{zheng2023codegeex}, MBXP~\citep{athiwaratkun2023multi}, MCEval~\citep{chaiMcEvalMassivelyMultilingual2024}, and ClassEval~\citep{duClassEvalManuallyCraftedBenchmark2023}. These datasets are selected for their diverse programming languages, task complexities, and availability of solution code segments, offering a comprehensive foundation for code generation tasks.
Compared to other datasets like DS-1000~\citep{lai2023ds} or APPS~\citep{hendrycks2measuring} and HumanEval~\citep{chenEvaluatingLargeLanguage2021}, which often focus on isolated, single-language problems, the selected datasets in Flow2Code offer a better balance of multilingual coverage and task variety. Furthermore, the inclusion of class-level tasks from ClassEval allows for a more holistic evaluation, challenging models with more complex code generation.

Moreover, only those problems with official solution code segments are kept. The availability of official solutions ensures the reliability and consistency of the flowchart generation process, enabling us to evaluate flowchart-based code generation tasks effectively.

Finally, the selected data is obtained. These selections ensure a rich and varied dataset that tests code generation models across a wide range of programming languages and task complexities.

\subsection{Flowchart Construction}\label{DC:2}

In the Flow2Code dataset construction process, a crucial step is transforming the raw data from the original datasets into a format suitable for code generation. This transformation process relies on an automated pipeline designed to convert solution code segments into flowchart representations, facilitating their subsequent use in code generation tasks.

\subsubsection{Code Flowchart and UML Flowchart}
The construction progress of code flowchart and UML flowchart can be divided into three steps: format conversion, flowchart conversion, and human review.

\textbf{Format conversion}: 
The first step in this transformation process involves converting the solution code segments into code files. The solution code segments contain code in various programming languages, covering diverse task complexities.

\textbf{Flowchart conversion}:
As shown in Figure~\ref{figure:gen-flow}, this work uses Visustin software\footnote{\url{https://www.aivosto.com/visustin.html}}, 
a widely-used tool for code-to-flowchart conversion, to generate code flowcharts and UML flowcharts from these code files. 
Visustin processes the code into flowcharts that visually represent the control flow, function calls, conditionals, and other programmatic steps, providing a clear and intuitive view of the underlying logic.

\textbf{Human review}:
To ensure the correctness and fidelity of the generated flowcharts, human evaluation is conducted. 
Five evaluators with master's degrees in computer science who have over four years of practical programming experience are employed. 
Each evaluator is proficient in multiple programming languages and software engineering practices.
The evaluation is conducted in two phases. 
For code and UML flowcharts, the evaluators verify the logical correctness of the generated flowcharts against the source code using the following criteria: \textbf{Logical consistency}: Do flowchart execution paths, conditions, and loops exactly reflect the original code? \textbf{Completeness}: Are there missing or extraneous elements that alter the intended program logic? \textbf{Semantic accuracy}: Are variable assignments, function calls, and control statements correctly captured?
Each instance is scored in binary form: \textbf{1}: The flowchart is logically and semantically correct and complete. \textbf{0}: The flowchart contains critical inaccuracies or omissions.
All instances are independently double-blind reviewed by two evaluators. 
In case of disagreement, the third expert adjudicates. The Krippendorff’s Alpha~\citep{krippendorff2011computing} is employed to assess inter-annotator agreement, achieving a reliability score of \textit{α = 0.88}. This high agreement ensures the trustworthiness of the manual verification.

 


\subsubsection{Pseudocode Flowchart}
Pseudocode flowcharts provide a human-readable and language-agnostic description of the program's logic, which complements the visual structure of code and UML flowcharts. This combination of textual pseudocode with visual flowchart representations enables more comprehensive and interpretable model evaluations. 
It helps to bridge the gap between abstract flow control and practical code generation by providing an additional level of clarity that may not be captured fully by flowcharts or UML diagrams alone. 

A particularly important step in the dataset construction process is the conversion of the raw DOT files into a pseudocode flowchart. 
The pseudocode flowchart construction can be divided into six steps:  
format conversion, Dot code generation, GPT-4o transformation, Gemini-2.0 check, human review, and automatic flowchart conversion.

\textbf{Format conversion}:
The format conversion is conducted the same as the format conversion in code flowchart and UML flowchart conversion.

\textbf{Dot code generation}:
Furthermore, Visustin also converts the flowcharts into raw DOT files, which are descriptions of graphs in a specific format used by the Graphviz visualization tool~\citep{ellson2004graphviz}.
In these DOT files, each node corresponds to a specific step or operation in the code’s flow, such as a variable assignment or a loop, while the edges between nodes represent the relationships between these steps. The DOT format provides a flexible, machine-readable structure that can be used for various types of graph visualizations.

\textbf{GPT-4o transformation}: 
The GPT model is given the task of converting each node label in the DOT file into a natural language description. 
The prompt is carefully designed to ensure the labels clearly describe the logic of each node in plain English (the prompt is in Appendix Figure~\ref{figure:Prompt for DOT Code Transformation}). 
As shown in Figure~\ref{figure:gen-flow}, a label like ``a, b = b, a \% b'' might be transformed into ``Assign `b' to `a', and the remainder of `a' divided by `b' to `b'.'' a conditional like ``b ?'' might be converted into ``check if `b' is equal to zero.'' 

\textbf{Gemini check}:
Since GPT-4o is used to generate the pseudocode DOT files, using the same model to also validate the results could introduce the risk of potential biases or errors in the verification process. 
To mitigate this concern, Gemini-2.0 is tasked with determining whether the content of the generated DOT files has been correctly converted into natural language descriptions by GPT-4o.
The prompt used for Gemini-2.0's verification is detailed in Appendix Figure~\ref{figure:Prompt for Dot Code Check}.
If Gemini-2.0 determines that the content is accurate, the corresponding data is retained in the dataset. 
If Gemin-2.0 identifies errors in an instance, the instance is re-generated by GPT-4o.


To maintain the integrity of the dataset, this re-generation process is repeated up to five times. 
If the data still fails to meet the accuracy criteria after five attempts, 
it is excluded from the final dataset to prevent erroneous or unreliable data from affecting the benchmark.
This approach guarantees that only high-quality, accurately converted data is included in Flow2Code, ensuring the reliability of the dataset for subsequent code generation evaluations. 

\textbf{Human review}: 
In verifying the pseudocode flowcharts derived from natural language transformations of DOT representations, evaluators conduct a structured assessment based on:
\textbf{Semantic accuracy}: Whether natural language node labels precisely match the intended semantics of the original DOT nodes.
\textbf{Clarity}: Whether the labels are concise and easily interpretable.
\textbf{Consistency}: Whether the number and structure of descriptions accurately correspond to the graph structure.
Scoring follows a binary scheme:
\textbf{1}: Label is accurate, clear, and consistent with the original structure.
\textbf{0}: Label is ambiguous, incorrect, or inconsistent.
Each pseudocode flowchart is reviewed independently by at least two evaluators, with a third reviewer resolving conflicts. 
As with the other flowcharts, the scoring achieves Krippendorff’s Alpha of \textit{0.88}.

\textbf{Automatic flowchart conversion}:
Finally, Graphviz is used to automatically convert the corresponding pseudocode flowcharts from the generated DOT files.

\begin{figure}[t]
  \includegraphics[width=\columnwidth]{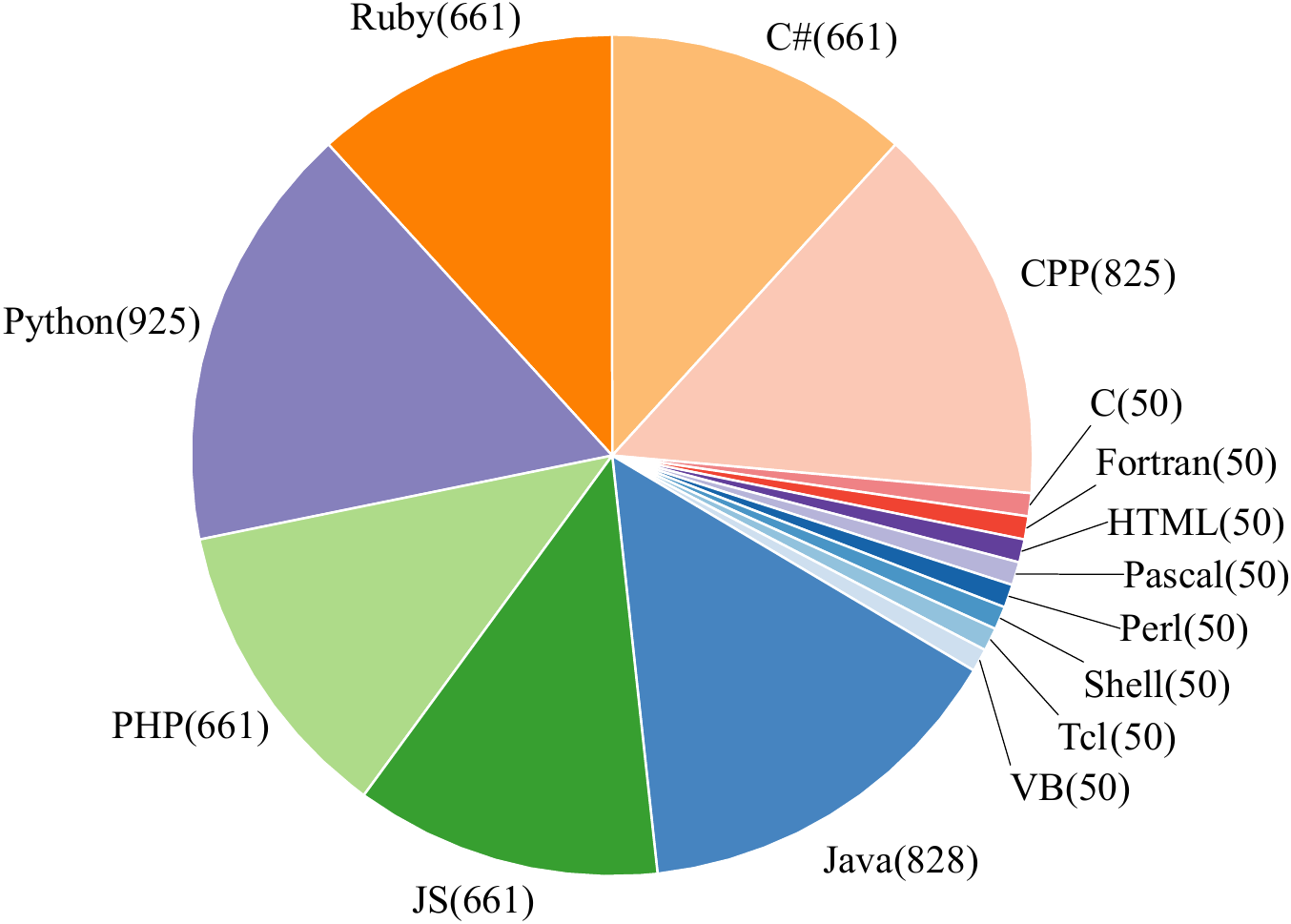}
  \caption{The number and proportion of each programming language in the Flow2Code dataset.}
  \label{fig:lang_num}
\end{figure}


\subsection{Dataset Analysis}\label{DC:3}
\subsubsection{Data Statistics}

Through this transformation process, the Flow2Code dataset ultimately includes three types of flowcharts—code flowcharts, UML flowcharts, and pseudocode flowcharts—and contains 15 different programming languages, 5,622 code segments, and 16,866 flowcharts, as shown in Table~\ref{table: dataset statistics}.
The distribution of samples across these languages reflects the inherent characteristics of the original datasets, where some languages are more represented than others, as illustrated in Figure~\ref{fig:lang_num}.
We offer a well-rounded multimodal approach that better tests and pushes the code generation capabilities of MLLMs. This ensures that models are evaluated in a more thorough, balanced manner, improving their ability to generate accurate and meaningful code from diverse input formats.

To support fine-tuning and evaluation experiments, the Flow2Code dataset is divided into training and test sets using a 9:1 split ratio. 
This partitioning is conducted within each language subset of the original source datasets (e.g., HumanEval-X, MBXP, McEval, and ClassEval). 
For instance, in HumanEval-X, which contains code samples across multiple programming languages, we select 10\% of the instances from each language as the test set and retain the remaining 90\% for training. This approach ensures proportional representation across all languages and prevents the exclusive use of any source dataset as the test set. Such balanced partitioning enhances the fairness and generalizability of our experimental evaluations.

\begin{table}[]
\small
\centering
\begin{tabular}{@{}lr@{}}
\toprule
\# of Programming Languages & 15 \\
\# of Code Segments & 5,622 \\
Avg. \# of Characters in Code Segments & 315.93 \\
Max. \# of Characters in Code Segments & 3914 \\
Min. \# of Characters in Code Segments & 10 \\ \midrule
\# of Code Flowchart & 5,622 \\
\# of UML Flowchart & 5,622 \\
\# of Pseudocode Flowchart & 5,622 \\
Avg. \# of Units in Flowcharts & 11.44 \\
Max. \# of Units in Flowcharts & 84 \\
Min. \# of Units in Flowcharts & 3 \\ \bottomrule
\end{tabular}%

\caption{Statistics of the Flow2Code dataset.}
\label{table: dataset statistics}
\end{table}

\subsubsection{Data Quality}
To ensure the quality of the Flow2Code dataset, we conduct a two-step human evaluation process, following the approach of \citet{liu2020towards}. 
Five evaluators, all graduate students with master's degrees in computer science and over four years of programming experience, are selected to carry out the evaluation.

The evaluation includes two steps:

\textbf{Code and UML Flowchart Verification}: Evaluators checked whether the flowcharts accurately represented the solution code segment’s logic. 
Each correct conversion is assigned a score of ``1'' and an incorrect one ``0''.

\textbf{Pseudocode Flowchart Verification}: Evaluators judge whether GPT correctly transforms the DOT file labels into natural language descriptions. 
Each correct conversion is assigned a score of ``1'' and an incorrect one ``0''.

We randomly sampled 100 instances from each flowchart type for evaluation. 
The average score across all evaluations is 0.94, indicating that the dataset contains high-quality flowcharts suitable for code generation tasks. 
This process ensures the reliability of Flow2Code for further research and evaluation.

\begin{table*}[t]
\resizebox{\textwidth}{!}{%
\begin{tabular}{@{\hskip 0.2cm}lcccccccccccccc@{}}
\toprule
\multicolumn{1}{l|}{\multirow{2}{*}{Model}} & \multicolumn{1}{c|}{ClassEval (100)} & \multicolumn{4}{c|}{HumanEval-X (164)} & \multicolumn{7}{c|}{MBXP (611)} & \multicolumn{2}{c}{McEval (50)} \\ \cmidrule(l){2-15} 
\multicolumn{1}{l|}{} & \multicolumn{1}{c|}{Python} & CPP & Java & JS & \multicolumn{1}{c|}{Python} & CPP & C\# & Java & JS & PHP & Python & \multicolumn{1}{c|}{Ruby} & C & C\# \\ \midrule
Claude-3.5-Sonnet & 26.00 & 32.32 & 48.78 & 39.02 & 49.39 & 43.86 & 51.23 & 50.90 & 51.06 & \textbf{100.00} & 57.77 & 49.26 & 48.00 & 70.00 \\
DeepSeek-VL2 & 1.00 & 1.83 & 11.59 & 29.88 & 40.24 & 12.77 & 11.29 & 9.00 & 39.12 & 85.27 & 42.06 & 45.01 & 28.00 & 36.00 \\
Gemini-2.0 & \textbf{70.00} & \textbf{85.37} & \textbf{90.24} & \textbf{87.20} & \textbf{87.80} & \textbf{90.67} & \textbf{82.98} & \textbf{89.20} & \textbf{92.14} & 91.49 & \textbf{94.27} & \textbf{94.11} & \textbf{78.00} & \textbf{96.00} \\
GLM-4V-plus & 18.00 & 54.27 & 60.37 & 78.66 & 72.56 & 76.60 & 68.58 & 75.12 & 81.83 & 98.36 & 79.38 & 84.78 & 68.00 & 72.00 \\
GPT-4o & {\ul 47.00} & 43.90 & {\ul 82.93} & {\ul 85.98} & 82.93 & 74.14 & 67.92 & 71.85 & {\ul 85.76} & {\ul 99.67} & {\ul 87.23} & 89.20 & {\ul 76.00} & {\ul 90.00} \\
Intern-VL2.5-8B-MPO & 14.00 & 12.80 & 21.95 & 56.70 & 59.75 & 31.75 & 14.40 & 21.44 & 60.07 & 99.51 & 68.74 & 72.01 & 38.00 & 30.00 \\
Intern-VL2.5-78B-MPO & 35.00 & 53.05 & 68.29 & 79.27 & {\ul 84.15} & 72.01 & 71.85 & 58.10 & 81.01 & 97.87 & 85.11 & {\ul 89.85} & 60.00 & 78.00 \\
LLaVA-OneVision-7B & 0.20 & 0.00 & 6.71 & 3.66 & 10.37 & 0.00 & 0.00 & 1.47 & 18.99 & 65.47 & 9.33 & 13.75 & 2.00 & 2.00 \\
LLaVA-OneVision-72B & 7.00 & 9.15 & 26.22 & 34.15 & 42.07 & 23.24 & 13.75 & 18.99 & 46.64 & \textbf{100.00} & 45.34 & 52.05 & 26.00 & 36.00 \\
MiniCPM-V-2\_6 & 2.00 & 1.22 & 14.02 & 21.34 & 34.76 & 5.56 & 10.47 & 3.44 & 45.50 & 83.47 & 42.23 & 54.17 & 4.00 & 2.00 \\
Qwen2-VL-72B & 41.00 & {\ul 54.88} & 74.39 & 74.39 & 79.27 & 76.60 & 65.47 & 70.21 & 84.94 & \textbf{100.00} & 83.63 & 88.38 & 66.00 & 84.00 \\
Qwen2-VL-7B & 14.00 & 1.83 & 38.41 & 50.00 & 64.63 & 10.15 & 3.60 & 14.89 & 54.99 & 66.28 & 54.34 & 28.48 & 32.00 & 44.00 \\
Qwen2-VL-7B-FT & 25.00 & 49.39 & 75.61 & 71.34 & 79.88 & {\ul 84.12} & {\ul 81.34} & {\ul 82.65} & 81.83 & 86.42 & 78.56 & 83.63 & 64.00 & 76.00 \\ \midrule
Variance & 4.10 & 9.52 & 9.23 & 7.49 & 5.67 & 11.55 & 11.28 & 11.14 & 5.09 & 1.47 & 5.93 & 6.65 & 6.36 & 9.92 \\
Avg & 24.09 & 34.80 & 50.57 & 56.88 & 62.46 & 48.99 & 44.90 & 46.61 & 64.94 & 89.77 & 65.38 & 66.40 & 46.57 & 56.86 \\ \midrule
\multicolumn{1}{l|}{\multirow{2}{*}{Model}} & \multicolumn{14}{c}{McEval (50)} \\ \cmidrule(l){2-15} 
\multicolumn{1}{l|}{} & \multicolumn{1}{c|}{CPP} & \multicolumn{1}{c|}{Fortran} & \multicolumn{1}{c|}{HTML} & \multicolumn{1}{c|}{Java} & \multicolumn{1}{c|}{JS} & \multicolumn{1}{c|}{Pascal} & \multicolumn{1}{c|}{Perl} & \multicolumn{1}{c|}{PHP} & \multicolumn{1}{c|}{Python} & \multicolumn{1}{c|}{Ruby} & \multicolumn{1}{c|}{Shell} & \multicolumn{1}{c|}{Tcl} & \multicolumn{1}{c|}{VB} & Avg \\ \midrule
Claude-3.5-Sonnet & 48.00 & 32.00 & 12.00 & 32.08 & 50.00 & 52.00 & 54.00 & 62.00 & 60.00 & 68.00 & 48.00 & 56.00 & 70.00 & 50.43 \\
DeepSeek-VL2 & 26.00 & 12.00 & 6.00 & 33.96 & 34.00 & 4.00 & 32.00 & 38.00 & 46.00 & 32.00 & 12.00 & 10.00 & 32.00 & 26.53 \\
Gemini-2.0 & \textbf{86.00} & 24.00 & 52.00 & {\ul 35.84} & \textbf{92.00} & \textbf{70.00} & \textbf{92.00} & \textbf{94.00} & 84.00 & \textbf{90.00} & 48.00 & \textbf{82.00} & {\ul 86.00} & \textbf{80.48} \\
GLM-4V-plus & 64.00 & 46.00 & 30.00 & 30.19 & 68.00 & 48.00 & 52.00 & 86.00 & 74.00 & 78.00 & 42.00 & 54.00 & 52.00 & 63.49 \\
GPT-4o & 64.00 & {\ul 68.00} & {\ul 60.00} & \textbf{35.85} & {\ul 84.00} & {\ul 66.00} & 74.00 & {\ul 92.00} & \textbf{94.00} & {\ul 82.00} & \textbf{66.00} & {\ul 74.00} & \textbf{92.00} & {\ul 75.43} \\
Intern-VL2.5-8B-MPO & 44.00 & 20.00 & 14.00 & 24.53 & 38.00 & 24.00 & 38.00 & 32.00 & 48.00 & 52.00 & 14.00 & 22.00 & 36.00 & 37.34 \\
Intern-VL2.5-78B-MPO & 58.00 & 60.00 & \textbf{66.00} & 26.42 & 68.00 & 46.00 & {\ul 76.00} & 60.00 & 82.00 & 80.00 & 50.00 & 66.00 & 74.00 & 67.71 \\
LLaVA-OneVision-7B & 4.00 & 0.00 & 0.00 & 5.66 & 10.00 & 4.00 & 8.00 & 20.00 & 68.00 & 50.00 & 2.00 & 4.00 & 0.00 & 11.47 \\
LLaVA-OneVision-72B & 24.00 & 4.00 & 2.00 & 33.96 & 30.00 & 24.00 & 30.00 & 24.00 & 42.00 & 60.00 & 12.00 & 20.00 & 48.00 & 30.91 \\
MiniCPM-V-2\_6 & 10.00 & 0.00 & 0.00 & 1.89 & 14.00 & 0.00 & 28.00 & 12.00 & 34.00 & 48.00 & 8.00 & 4.00 & 0.00 & 18.31 \\
Qwen2-VL-72B & 66.00 & \textbf{72.00} & 40.00 & 30.19 & 74.00 & 50.00 & 74.00 & 78.00 & {\ul 84.00} & 78.00 & {\ul 64.00} & 66.00 & 80.00 & 70.35 \\
Qwen2-VL-7B & 42.00 & 22.00 & 4.00 & 26.42 & 48.00 & 18.00 & 24.00 & 56.00 & 46.00 & 64.00 & 32.00 & 20.00 & 52.00 & 35.05 \\
Qwen2-VL-7B-FT & {\ul 70.00} & 36.00 & 20.00 & 20.75 & 68.00 & 26.00 & 56.00 & 76.00 & 64.00 & 70.00 & 46.00 & 56.00 & 68.00 & 62.85 \\ \midrule
Variance & 5.81 & 6.32 & 5.25 & 1.09 & 6.31 & 5.60 & 5.73 & 7.59 & 3.49 & 2.56 & 5.32 & 8.05 & 8.58 & 5.30 \\
Avg & 47.71 & 32.57 & 24.14 & 26.01 & 52.86 & 35.00 & 49.43 & 57.00 & 64.29 & 65.86 & 36.57 & 43.29 & 54.71 & 49.95 \\ \bottomrule
\end{tabular}%
}
  \caption{\label{flowchart pass@1 all}
    Pass@1 results for 13 LLMs on the code flowchart generation task in the Flow2Code benchmarks. 
    ClassEval, HumanEval-X, MBXP, and McEval represent the code source of the instances in Flow2chart, respectively. The results in bold are the optimal results, while the underlined results represent the suboptimal results.
    The results are represented in percentage (\%).
  }
  \label{table:pass1_all}
\end{table*}

\section{Experiments and Results}
This section first presents the experimental setting (Section~\ref{ER:1}) and evaluation metrics (Section~\ref{ER:2}), followed by the model baselines (Section~\ref{ER:3}) and experimental results (Section~\ref{ER:4}).

\subsection{Experimental Setting}\label{ER:1}
\textbf{Implementation Details}: In this work, the local model deployment frameworks LMDeploy (version 0.6.5) and SGLang (version 0.4.1.post4) are used, with model fine-tuning conducted via the Llama-Factory framework (version 0.9.2.dev0). 

\textbf{Computing Platform}: Experiments are carried out on a server featuring dual Intel Xeon Gold 5320 CPUs, 377 GB RAM, and eight NVIDIA A100 GPUs, running Ubuntu 20.04.6 LTS. The setup provides a robust environment for efficient model training and deployment.

\textbf{Fine-tuning Details}: Qwen2-VL-7B is fine-tuned on the Flow2Code dataset using Low-Rank Adaptation (LoRA)~\citep{hulora}, chosen for its efficient adaptation of large models to multimodal tasks. 
The fine-tuning employs a learning rate of 5.0e-5, a batch size of 4, and 1 training epoch with a cosine learning rate scheduler. 
The AdamW~\citep{loshchilov2017decoupled} optimizer is used with gradient accumulation steps of 8 to ensure stability (Details can be found in the Appendix Section~\ref{Fine-tuning Setting}).

\begin{figure*}[t]
  \includegraphics[width=1.0\linewidth]{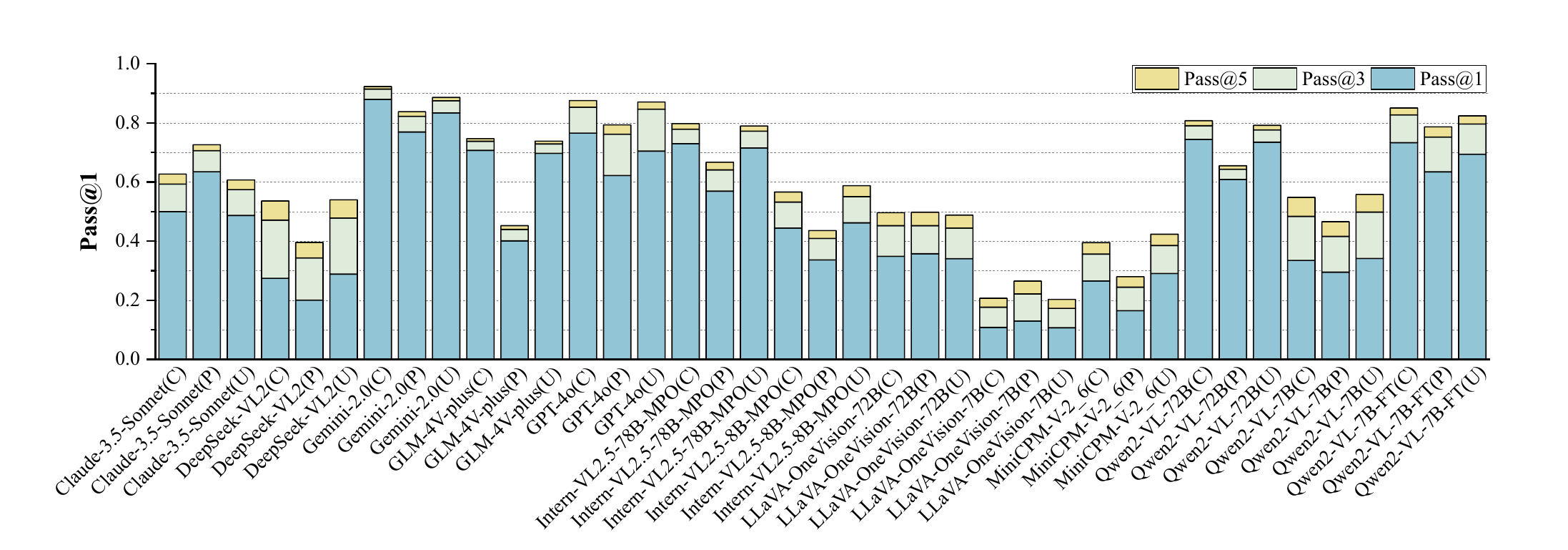}
  \caption {Stacked diagram of Pass@1, 3, and 5 of all the evaluation models on the benchmark. The suffixes of the model names represent different flowchart types: C represents code flowchart, P represents pseudocode flowchart, and U represents UML flowchart.}
  \label{figure:Stacked-pass}
\end{figure*}

\subsection{Evaluation Metrics}\label{ER:2}
The evaluation uses test samples from the four core datasets selected for Flow2Code. The testing procedure follows the execution-based evaluation method, where candidate code is executed against a set of test cases, and success is determined by passing those tests. Unlike traditional n-gram evaluations, execution-based evaluation allows for functional correctness even if the generated solution differs in implementation from the reference solution. This flexibility is crucial for code generation tasks, as it accommodates different code styles and approaches that still achieve the desired functionality.

Following~\citet{athiwaratkun2023multi}, we use Pass@\textit{k} scores~\citep{kulal2019spoc} with the unbiased estimate from \citet{chenEvaluatingLargeLanguage2021} as the evaluation metrics, where a task is deemed successful if any of the top \textit{k} samples are correct. 
For the evaluation, \textit{k} is set as 1, 3, and 5, as the key metrics, which provide a comprehensive view of the model's ability to generate correct code given flowchart.


\subsection{Baselines}\label{ER:3}
For the experiments, we evaluate a range of MMLLMs with a focus on their ability to generate code. 
The specific information of the model is shown in Appendix Table~\ref{table: baseline details}.

Based on the research of~\citet{shiri-etal-2024-empirical},~\citet{das-etal-2024-exams}, and~\citet{ai-etal-2024-advancement}, The following LLMs are evaluated:

\textbf{Claude-3.5-Sonnet}~\citep{anthropic2024claude} is known for its strong performance in reasoning, math, and coding tasks, particularly with multimodal inputs.

\textbf{DeepSeek-VL2}~\citep{wuDeepSeekVL2MixtureofExpertsVisionLanguage2024} is a Mixture-of-Experts model designed for visual understanding, ideal for tasks involving complex visual inputs.

\textbf{Gemini-2.0-Flash-Exp}~\citep{IntroducingGemini202024} offers low-latency, high-performance multimodal capabilities, particularly excelling in video understanding.

\textbf{GLM-4V-Plus}~\citep{glmChatGLMFamilyLarge2024} specializes in image and video recognition, making it suitable for tasks requiring advanced visual comprehension.

\textbf{GPT-4o}~\citep{openaiGPT4oSystemCard2024} is a leading multimodal model capable of processing text, images, and audio, with strong code generation abilities.

\textbf{InternVL2\_5-MPO (8B and 78B)}~\citep{chenInternVLScaling2024} is optimized for multimodal reasoning tasks, this model excels in complex, multimodal code generation scenarios.

\textbf{LLaVA-OneVision-Qwen2-OV (7B and 72B)}~\citep{li2025llavaonevision} is known for strong performance across a range of visual tasks, including image and video understanding.

\textbf{MiniCPM-V-2\_6}~\citep{yao2024minicpm} is optimized for video understanding and multimodal reasoning with efficient processing of visual data.

\textbf{Qwen2-VL-Instruct (7B and 72B)}~\citep{wangQwen2VLEnhancingVisionLanguage2024} provides strong performance in understanding visual content across varying resolutions.

These models are selected based on their demonstrated capabilities in multimodal tasks and their potential for generating code from both visual and textual inputs.

\subsection{Experimental Results}\label{ER:4}


\begin{figure*}[t]
  \includegraphics[width=1\linewidth]{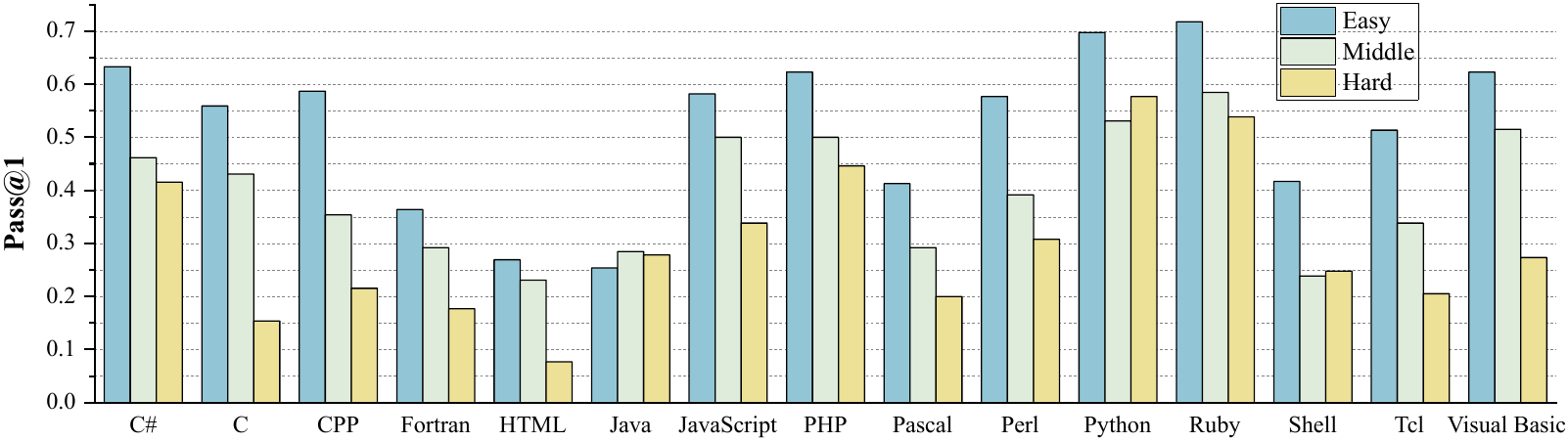}
  \caption {The average Pass@1 rate of the three difficulty levels of the samples in the Flow2Code's McEval part on the code flowchart.}
  \label{figure:pass-level}
\end{figure*}

As shown in Table~\ref{table:pass1_all} and Figure~\ref{figure:Stacked-pass}, the benchmark evaluation of 13 multimodal models across three flowchart types (code, UML, and pseudocode) reveals distinct performance patterns.

\textbf{Model Performance Disparity}: Gemini-2.0 consistently maintains a significant advantage across all three tasks, with an average pass rate improvement of 4.78\% over the next best models, i.e., GPT-4o. GPT-4o and Qwen2-VL-72B perform similarly on the code flowchart task, but their performance decreases by about 5\% and 10\% respectively on UML and pseudocode tasks. Smaller models (such as LLaVA-OneVision-7B and MINICPM-V-2\_6) exhibit overall weaker performance, with average pass rates only 14.3\%-22.35\% of those achieved by the top-performing Gemini-2.0.

\textbf{Programming Language Sensitivity}: The generation ability for PHP is notably strong, with three models achieving perfect scores. In contrast, the average pass rates for Fortran and HTML are 57.2\%-65.63\% lower compared to PHP. The disparity in generation capabilities between models is most pronounced for CPP, C\#, and JavaScript, which exhibit the largest variance in performance.
PHP, Python, and Ruby languages exhibit the least variation across models.

\textbf{Model Scale Effect}: Increasing the model size results in systematic improvements. For example, Qwen2-VL-72B shows a 103.8\% performance boost over its 7B version in the code flowchart task. Similarly, the Intern-VL2.5-78B version outperforms its 8B counterpart in the UML task by 75.1\%. 

\textbf{Flowchart Type Difference Results}: The code flowchart task shows the highest average pass rate, followed by the UML task, with the pseudocode task proving to be the most challenging, showing a decrease of 10.63\% and 7.48\% in average pass rates compared to the first two tasks. Some models exhibit task-specific performance, such as Claude-3.5-Sonnet, which performs exceptionally well in generating Tcl code for pseudocode tasks, but its ability to generate Fortran code decreases by 6\% compared to the code flowchart task.


\begin{figure}[t]
  \includegraphics[width=\columnwidth]{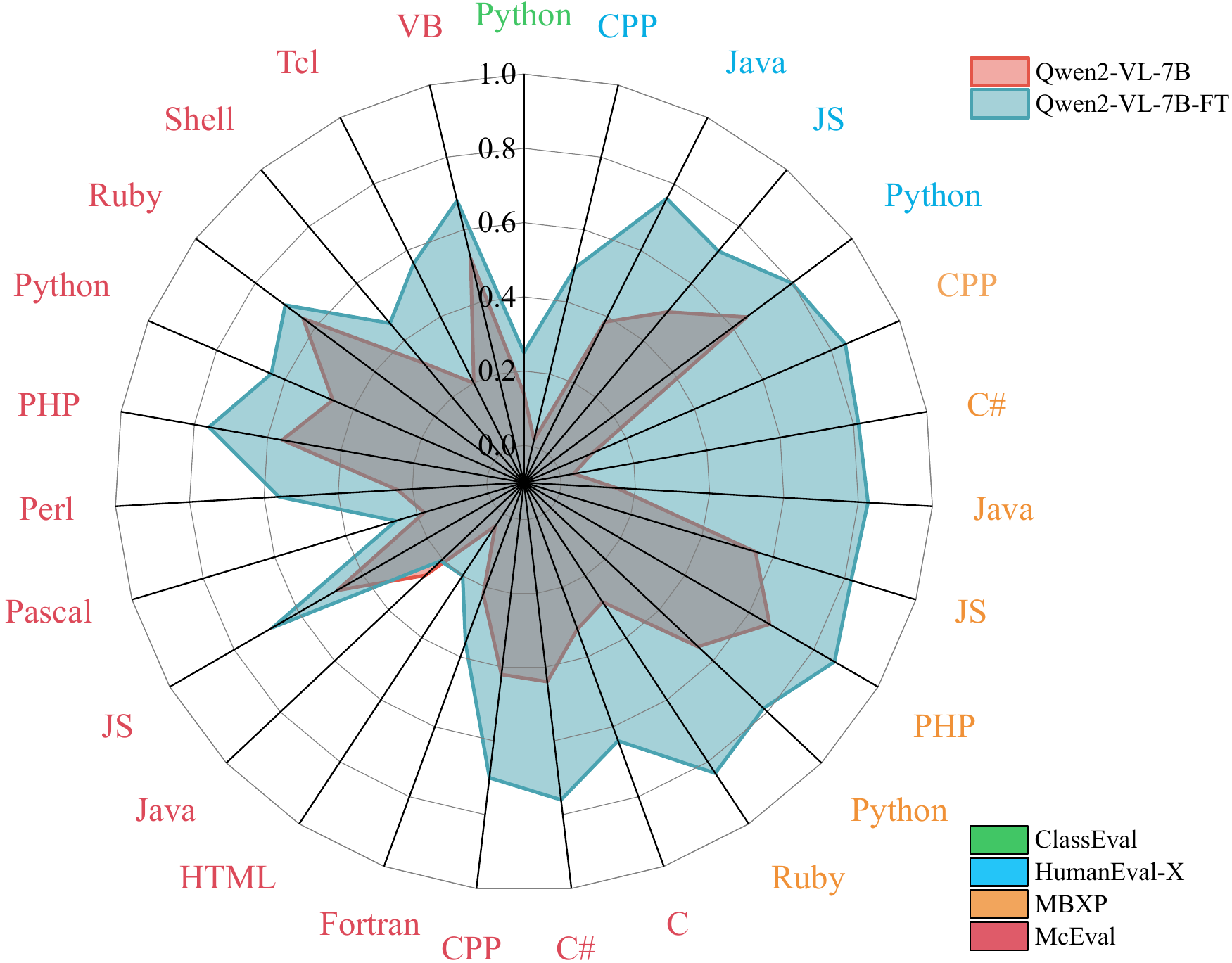}
  \caption{Average Pass@1 results for Qwen2-VL-7B and the fine-tuned Qwen2-VL-7B-FT (on the Flow2Code dataset) on code flowchart tasks. The dataset labels in the bottom-right corner of the image show which subclass corresponds to each dimension of the radar chart.}
  \label{fig:ft_radar}
\end{figure}

\textbf{Fine-Tuning Strategy Effectiveness}: As shown in Figure~\ref{fig:ft_radar}, after fine-tuning, significant performance improvements are observed in Qwen2-VL-7B-FT, with the most pronounced in object-oriented languages (C\#, Java) and web languages (JS, PHP), suggesting the fine-tuning data emphasized modern programming paradigms. The modest performance increase in legacy languages (Fortran, Pascal) and markup languages (HTML) suggests either data scarcity in these domains or architectural limitations in handling non-OOP paradigms.


\textbf{Code Generation with Different Difficulty Levels}: Figure~\ref{figure:pass-level} presents the average Pass@1 performance of 13 multimodal models across 15 programming languages, categorized by difficulty level (Easy, Middle, Hard). Several key observations can be made.

(1) A consistent trend across most languages and models is the inverse correlation between task difficulty and Pass@1. Performance is highest for ``Easy'' tasks, decreases for ``Middle'' tasks, and is lowest for ``Hard'' tasks. This is expected, as the complexity of the code generation task directly impacts the likelihood of generating a correct solution on the first attempt (Pass@1).

(2) There is significant variation in performance across different programming languages. For instance, models generally achieve higher Pass@1 scores in Python and Ruby, even at higher difficulty levels, compared to languages like HTML, Java, or Fortran. The superior performance of Python and Ruby may be attributed to a combination of factors, including the prevalence of these languages in training data, their relatively simpler syntax and higher-level abstractions, and the rich availability of libraries and corresponding descriptions within the dataset.

\section{Conclusion}
We introduce Flow2Code, a multimodal dataset combining flowcharts and code in 15 programming languages, designed to advance research in code generation. The dataset includes three flowchart types—code, UML, and pseudocode—providing a rich resource for training and evaluating MLLMs. We propose a new benchmark, Flow2Code, to assess MLLMs on code generation from flowchart inputs. Extensive evaluations of 13 MLLMs highlight their strengths and weaknesses, providing valuable insights for future research. 

\section*{Acknowledgments}
We want to thank the School of Computer Science and Technology and the Institute of AI Education at East China Normal University for providing the computational platform. We also thank the reviewers for their insightful comments.

\section*{Limitations}
While this study introduces the Flow2Code dataset and the Flow2Code benchmark as valuable resources for code generation, the limitations remain that should be addressed in future work. 
The current evaluation framework focuses on end-to-end code generation tasks but does not account for the potential complexities involved in real-world code maintenance tasks, such as debugging or refactoring. Extending the benchmark to include tasks beyond initial code generation, such as identifying and fixing bugs or optimizing existing code based on flowchart representations, could better reflect the broader utility of multimodal models in software development.

\section*{Ethics Statement}
We ensure that the Flow2Code dataset is constructed in full compliance with the terms of use of the original datasets (HumanEval-X, MBXP, MCEval, and ClassEval) and strictly respect the intellectual property rights of their authors. All code segments and flowchart transformations are derived from publicly available solutions or generated programmatically, with no inclusion of sensitive, proprietary, or personally identifiable information.

For human reviewers involved in the flowchart verification and pseudocode transformation processes, we guarantee fair treatment, including appropriate compensation and adherence to ethical labor practices. Reviewers participated voluntarily with full awareness of their tasks and associated requirements, and their contributions were anonymized to protect privacy.

The Flow2Code dataset focuses on flowchart-based code generation tasks and does not involve socially sensitive topics, biased content, or ethically controversial material. The generated flowcharts and code segments are purely algorithmic and logic-driven, posing no foreseeable risks of misuse or harmful societal consequences. All flowchart generation tools (Visustin, Graphviz) and AI models (GPT-4o, Gemini-2.0) were used in accordance with their respective licenses and API terms. The dataset will be released under open licenses that align with the original data sources’ policies.

\bibliography{custom, reference}

\appendix

\section{Flowchart Example}
Due to page limitations in the main text, it is not feasible to include relatively complex flowcharts. Therefore, a set of representative flowchart examples is provided in the appendix. These examples illustrate three different forms of flowchart representations for the same code segment.

\begin{figure*}[]
\centering
  \includegraphics[width=1\linewidth]{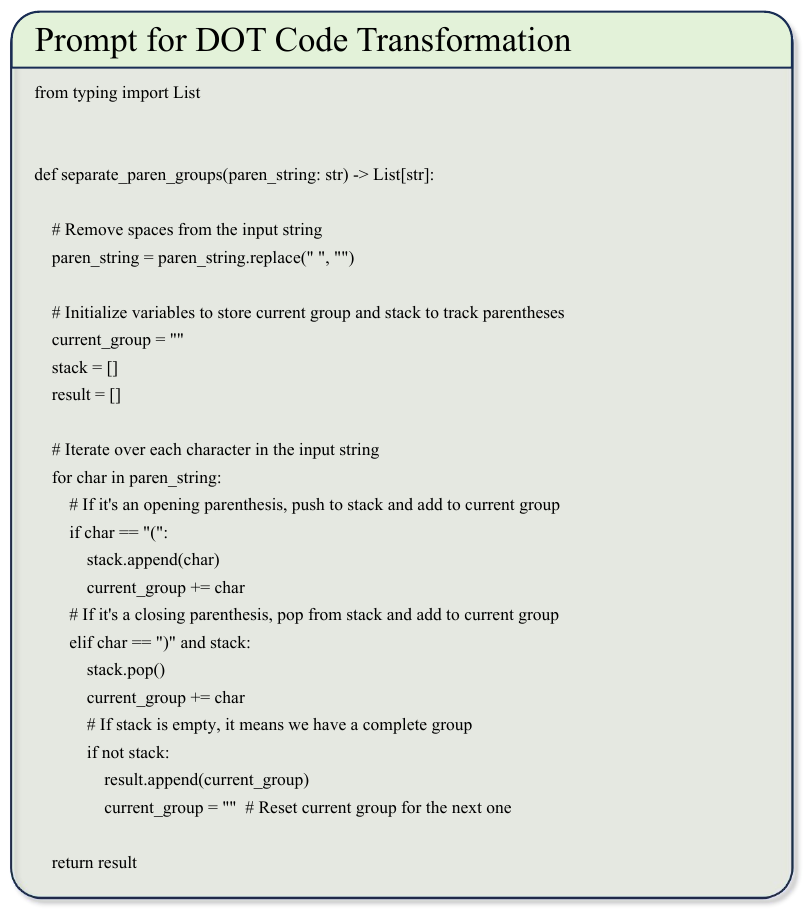}
  \caption {The code segment example.}
  \label{figure:code example}
\end{figure*}

\begin{figure*}[]
\centering
  \includegraphics[width=1\linewidth]{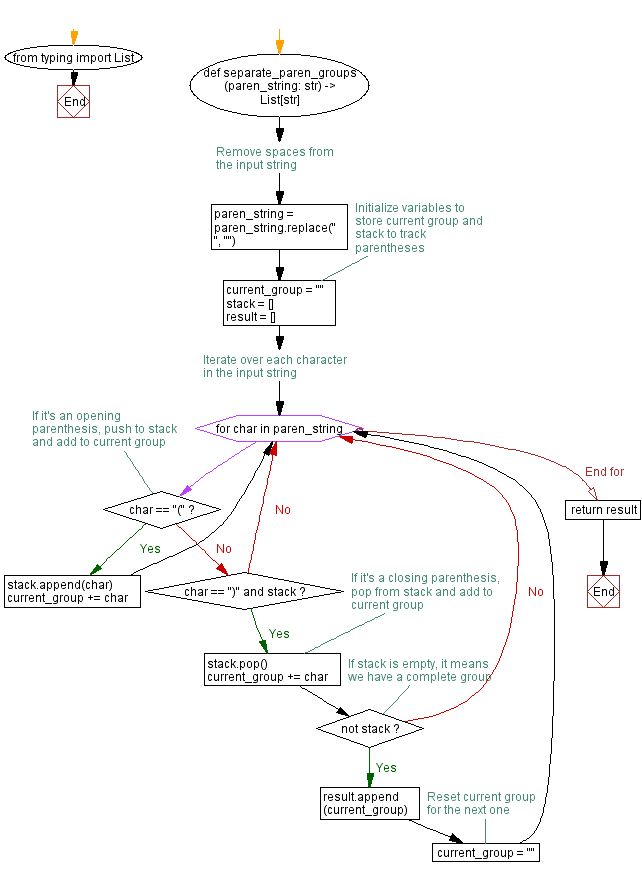}
  \caption {The code flowchart based on Figure~\ref{figure:code example} the code segment example.}
  \label{figure:code example flow}
\end{figure*}

\begin{figure*}[]
\centering
  \includegraphics[width=0.7\linewidth]{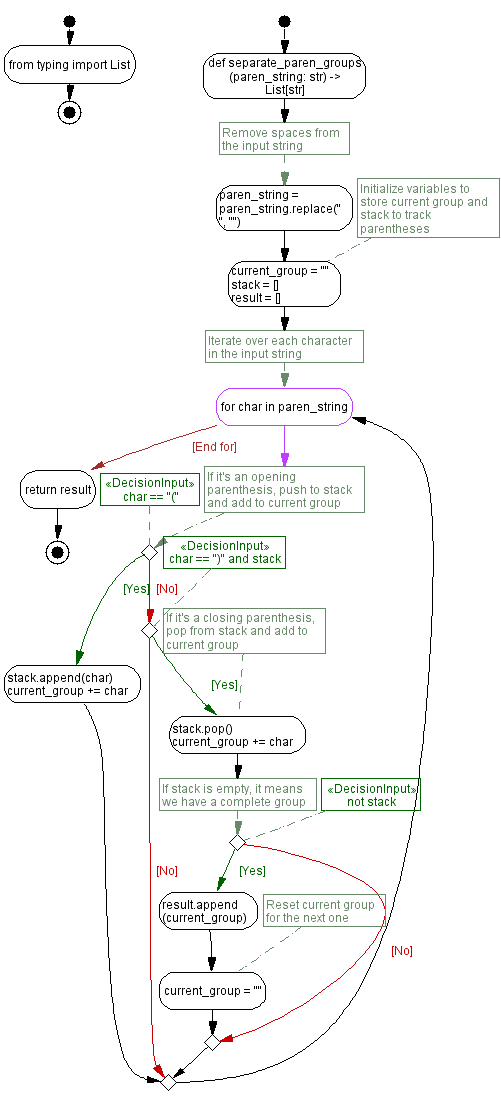}
  \caption {The UML flowchart based on Figure~\ref{figure:code example} the code segment example.}
  \label{figure:code example uml}
\end{figure*}

\begin{figure*}[]
\centering
  \includegraphics[width=1\linewidth]{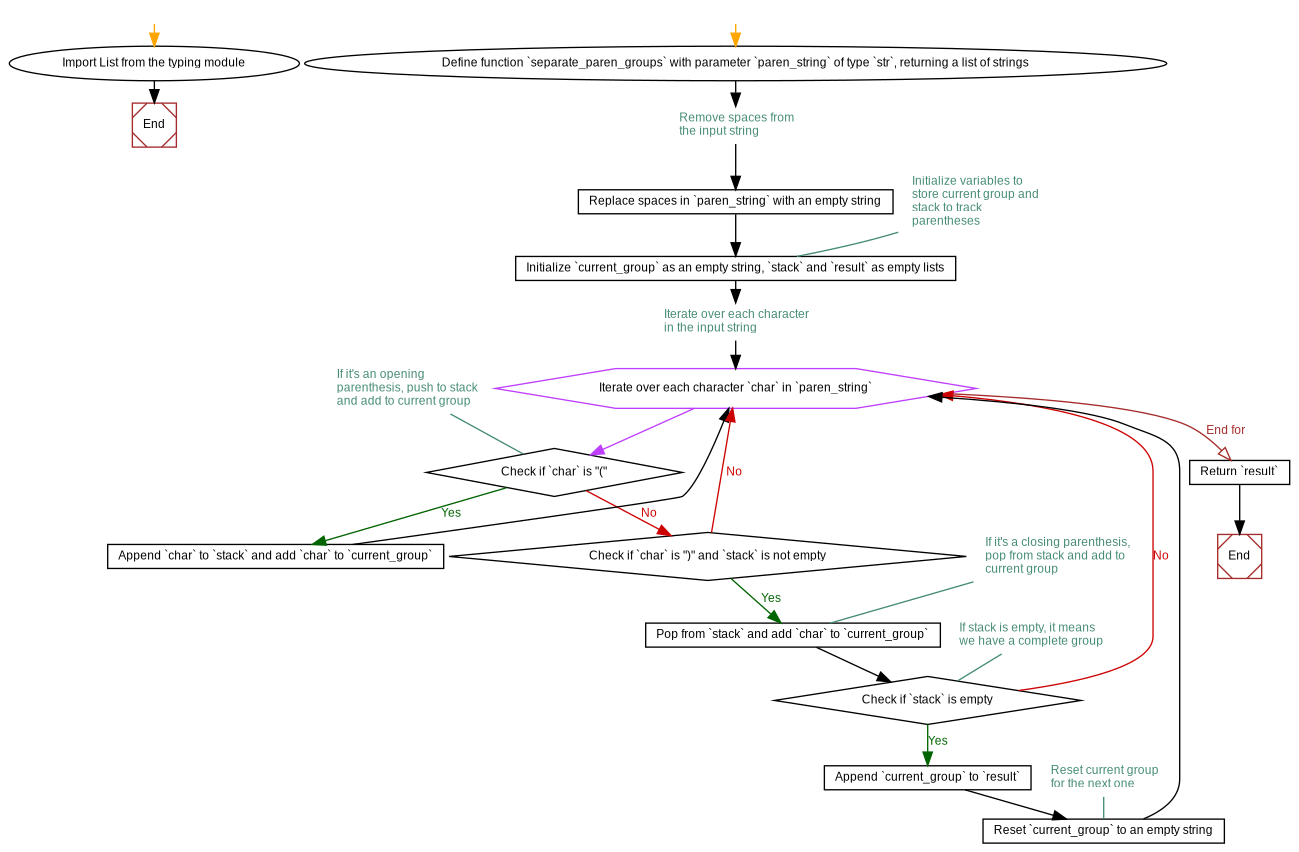}
  \caption {The pseudocode flowchart based on Figure~\ref{figure:code example} the code segment example.}
  \label{figure:code example pseudo}
\end{figure*}

\section{Prompts to Use}
\label{sec:appendix}
This section presents the prompts used during the flowchart generation process and Flow2Code benchmark evaluation, as well as the LLM message code templates.

\begin{figure*}[]
\centering
  \includegraphics[width=1\linewidth]{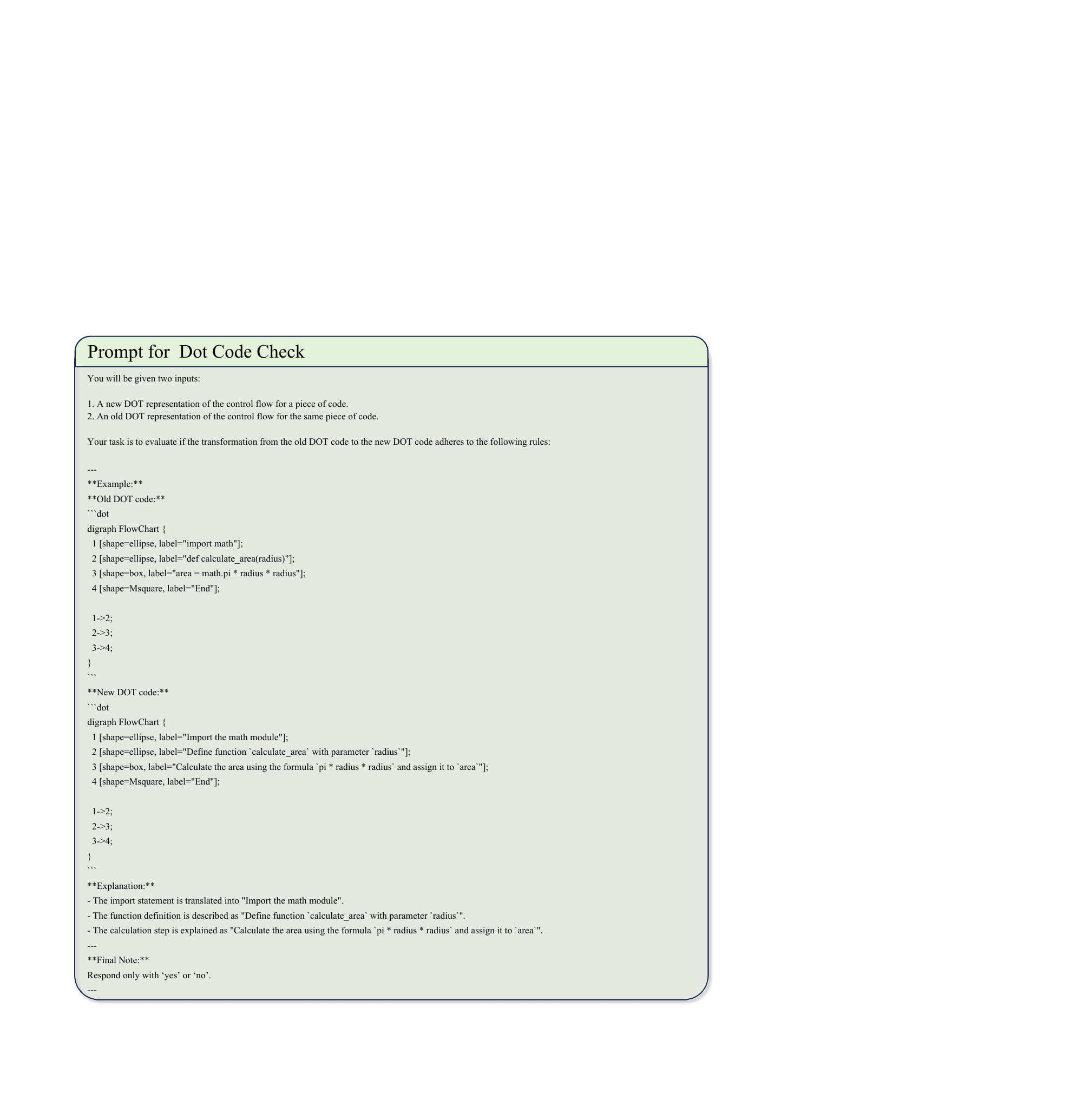}
  \caption {The prompt used by Gemini-2.0 to check the DOT code generated by GPT-4o.}
  \label{figure:Prompt for Dot Code Check}
\end{figure*}

\begin{figure*}[]
\centering
  \includegraphics[width=1\linewidth]{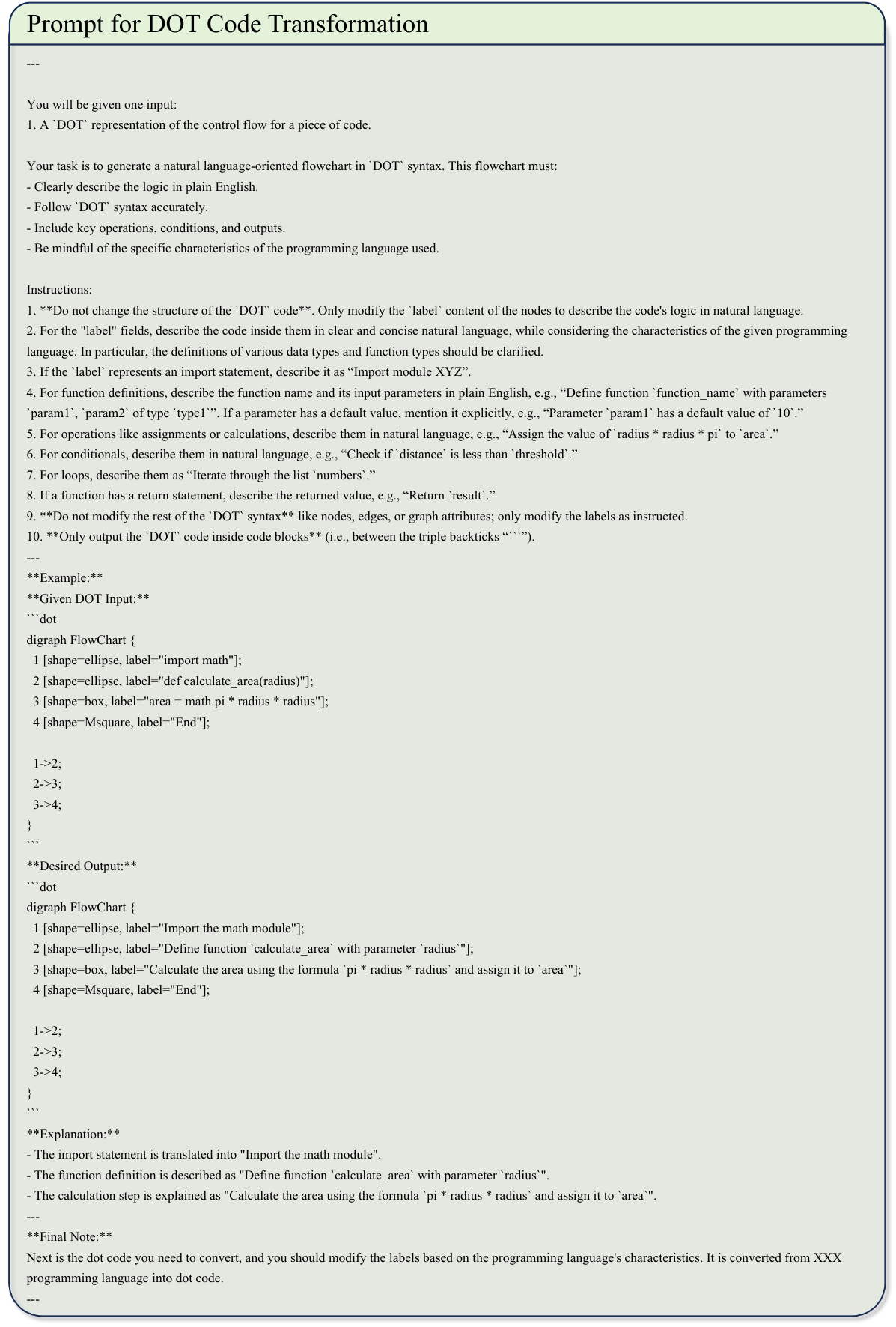}
  \caption {The prompt used by GPT-4o to convert DOT code into pseudocode DOT code.}
  \label{figure:Prompt for DOT Code Transformation}
\end{figure*}

\begin{figure*}[]
\centering
  \includegraphics[width=1\linewidth]{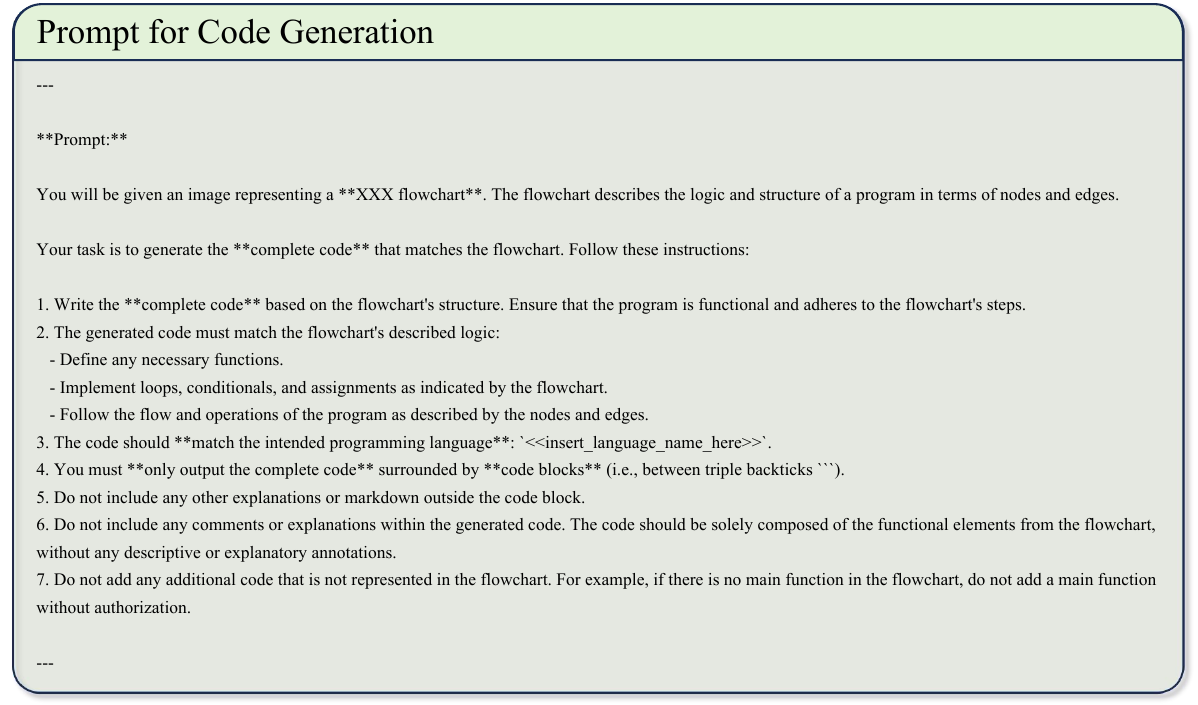}
  \caption {The prompt used to evaluate LLMs on flowchart-based code generation tasks using the Flow2Code benchmark.}
  \label{figure:prompt-code-gen}
\end{figure*}

\begin{figure*}[]
\centering
  \includegraphics[width=0.7\linewidth]{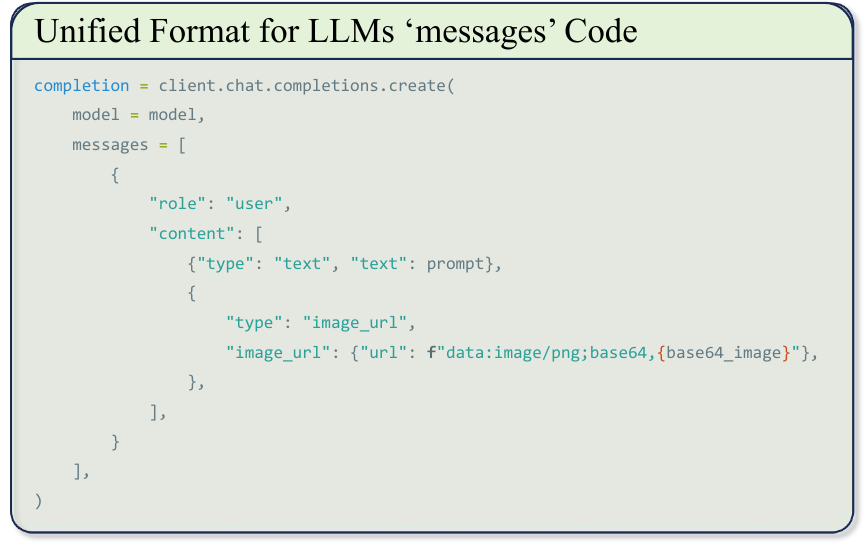}
  \caption{The LLM message code template used for evaluating large models on flowchart-based code generation tasks using the Flow2Code benchmark.}
  \label{fig:LLM message code template}
\end{figure*}





\section{Baselines Details}
This section provides detailed information about the 13 MLLMs used for the Flow2Code benchmark evaluation.

\begin{table*}[h]
\resizebox{\textwidth}{!}{%
\begin{tabular}{clccc}
\hline
\textbf{} & \textbf{Model} & \textbf{Language Model} & \textbf{Vision Model} & \textbf{time} \\ \hline
\multicolumn{1}{c|}{\multirow{4}{*}{\textbf{Closed-source}}} & Claude-3.5-Sonnet~\citep{anthropic2024claude} & - & - & 2024.10 \\
\multicolumn{1}{c|}{} & Gemini-2.0~\citep{IntroducingGemini202024} & - & - & 2024.12 \\
\multicolumn{1}{c|}{} & GLM-4V-plus~\citep{glmChatGLMFamilyLarge2024} & - & - & 2024.08 \\
\multicolumn{1}{c|}{} & GPT-4o~\citep{openaiGPT4oSystemCard2024} & - & - & 2024.05 \\ \hline
\multicolumn{1}{c|}{\multirow{8}{*}{\textbf{Open-source}}} & DeepSeek-VL2~\citep{wuDeepSeekVL2MixtureofExpertsVisionLanguage2024} & DeepSeekMoE-27B & SigLIP-400M & 2025.01 \\
\multicolumn{1}{c|}{} & Intern-VL2.5-8B-MPO~\citep{chenInternVLScaling2024} & InternLM2.5-7B & InternViT-300M-v2.5 & 2024.12 \\
\multicolumn{1}{c|}{} & Intern-VL2.5-78B-MPO~\citep{chenInternVLScaling2024} & Qwen-2.5-72B & InternViT-6B-v2.5 & 2024.12 \\
\multicolumn{1}{c|}{} & LLaVA-OneVision-7B~\citep{li2025llavaonevision} & Qwen2-7B & SigLIP-400M & 2024.09 \\
\multicolumn{1}{c|}{} & LLaVA-OneVision-72B~\citep{li2025llavaonevision} & Qwen2-72B & SigLIP-400M & 2024.09 \\
\multicolumn{1}{c|}{} & MiniCPM-V-2\_6~\citep{yao2024minicpm} & Qwen2-7B & SigLIP-400M & 2024.08 \\
\multicolumn{1}{c|}{} & Qwen2-VL-72B~\citep{wangQwen2VLEnhancingVisionLanguage2024} & Qwen2-72B & QwenViT & 2024.10 \\
\multicolumn{1}{c|}{} & Qwen2-VL-7B~\citep{wangQwen2VLEnhancingVisionLanguage2024} & Qwen2-7B & QwenViT & 2024.10 \\ \hline
\end{tabular}%
}
\caption{
Details of the baseline model.}
\label{table: baseline details}
\end{table*}

\section{Fine-tuning Setting}\label{Fine-tuning Setting}
The parameters used for fine-tuning are as follows:

bf16: true; cutoff\_len: 2048; ddp\_timeout: 180000000; do\_train: true; eval\_steps: 100; eval\_strategy: steps; finetuning\_type: lora; flash\_attn: auto; gradient\_accumulation\_steps: 8; learning\_rate: 5.0e-05; logging\_steps: 5; lora\_alpha: 16; lora\_dropout: 0; lora\_rank: 8; lora\_target: all; lr\_scheduler\_type: cosine; max\_grad\_norm: 1.0; max\_samples: 100000; num\_train\_epochs: 1.0; optim: adamw\_torch; packing: false; per\_device\_eval\_batch\_size: 4; per\_device\_train\_batch\_size: 4; plot\_loss: true; preprocessing\_num\_workers: 16; report\_to: none; save\_steps: 100; stage: sft; template: qwen2\_vl; trust\_remote\_code: true; val\_size: 0.1; warmup\_steps: 0.

\section{Additional Results}
This section presents the detailed results obtained from evaluating 13 MLLMs using the Flow2Code benchmark, including the specific Pass@1, Pass@3, and Pass@5 data for each of the three flowchart types.

\begin{table*}[]
\resizebox{\textwidth}{!}{%
\begin{tabular}{@{}lcccccccccccccc@{}}
\toprule
\multicolumn{1}{l|}{\multirow{2}{*}{Model}} & \multicolumn{1}{c|}{ClassEval (100)} & \multicolumn{4}{c|}{HumanEval-X (164)} & \multicolumn{7}{c|}{MBXP (611)} & \multicolumn{2}{c}{McEval (50)} \\ \cmidrule(l){2-15} 
\multicolumn{1}{l|}{} & \multicolumn{1}{c|}{Python} & CPP & Java & JS & \multicolumn{1}{c|}{Python} & CPP & C\# & Java & JS & PHP & Python & \multicolumn{1}{c|}{Ruby} & C & C\# \\ \midrule
Claude-3.5-Sonnet & 24.00 & 29.88 & 40.85 & 22.56 & 50.61 & 48.61 & 55.16 & 53.36 & 50.25 & \textbf{100.00} & 57.61 & 51.23 & 44.00 & 72.00 \\
DeepSeek-VL2 & 3.00 & 2.44 & 9.76 & 40.85 & 35.98 & 21.44 & 9.98 & 8.84 & 38.13 & 83.63 & 43.54 & 48.61 & 24.00 & 20.00 \\
Gemini-2.0 & \textbf{60.00} & \textbf{71.34} & \textbf{79.88} & \textbf{89.02} & \textbf{82.93} & \textbf{87.89} & \textbf{88.05} & {\ul 79.54} & \textbf{90.02} & 86.42 & \textbf{91.65} & \textbf{93.45} & \textbf{76.00} & {\ul 88.00} \\
GLM-4V-plus & 16.00 & {\ul 50.61} & 57.32 & 81.10 & 73.78 & 77.58 & 64.98 & 73.32 & 81.34 & 98.85 & 76.92 & 84.94 & 62.00 & 62.00 \\
GPT-4o & {\ul 43.00} & 37.80 & {\ul 73.17} & {\ul 80.49} & {\ul 79.27} & 60.56 & 58.27 & 69.07 & 76.92 & {\ul 99.18} & 82.16 & 85.43 & {\ul 66.00} & \textbf{90.00} \\
Intern-VL2.5-8B-MPO & 13.00 & 13.41 & 26.83 & 65.24 & 54.27 & 38.95 & 23.73 & 29.13 & 61.37 & 98.20 & 63.01 & 67.27 & 46.00 & 28.00 \\
Intern-VL2.5-78B-MPO & 37.00 & 42.68 & 64.63 & 82.32 & 75.00 & 72.01 & 71.69 & 62.03 & 80.20 & 99.84 & {\ul 83.96} & {\ul 87.07} & 60.00 & 80.00 \\
LLaVA-OneVision-7B & 0.20 & 0.00 & 3.05 & 9.15 & 5.49 & 0.00 & 0.00 & 3.11 & 19.15 & 62.36 & 10.80 & 15.22 & 4.00 & 0.00 \\
LLaVA-OneVision-72B & 9.00 & 7.93 & 23.78 & 39.02 & 37.20 & 20.46 & 13.91 & 19.64 & 43.21 & \textbf{100.00} & 45.01 & 49.92 & 28.00 & 32.00 \\
MiniCPM-V-2\_6 & 1.00 & 1.22 & 10.37 & 39.63 & 32.93 & 15.22 & 16.37 & 2.13 & 41.73 & 88.87 & 42.23 & 57.28 & 24.00 & 12.00 \\
Qwen2-VL-72B & 39.00 & 50.61 & {\ul 73.17} & 75.61 & 76.22 & 77.58 & 67.59 & 71.19 & {\ul 82.00} & 98.04 & 82.00 & 88.71 & 60.00 & 86.00 \\
Qwen2-VL-7B & 8.00 & 3.66 & 23.78 & 54.27 & 54.88 & 27.50 & 4.75 & 7.36 & 57.61 & 73.65 & 57.28 & 36.66 & 38.00 & 48.00 \\
Qwen2-VL-7B-FT & 20.00 & 48.78 & 71.34 & 60.98 & 65.24 & {\ul 81.51} & {\ul 82.65} & \textbf{81.18} & 81.51 & 84.78 & 71.52 & 83.47 & 56.00 & 0.00 \\ \midrule
\multicolumn{1}{l|}{\multirow{2}{*}{Model}} & \multicolumn{14}{c}{McEval (50)} \\ \cmidrule(l){2-15} 
\multicolumn{1}{l|}{} & \multicolumn{1}{c|}{CPP} & \multicolumn{1}{c|}{Fortran} & \multicolumn{1}{c|}{HTML} & \multicolumn{1}{c|}{Java} & \multicolumn{1}{c|}{JS} & \multicolumn{1}{c|}{Pascal} & \multicolumn{1}{c|}{Perl} & \multicolumn{1}{c|}{PHP} & \multicolumn{1}{c|}{Python} & \multicolumn{1}{c|}{Ruby} & \multicolumn{1}{c|}{Shell} & \multicolumn{1}{c|}{Tcl} & \multicolumn{1}{c|}{VB} & Avg \\ \midrule
Claude-3.5-Sonnet & 46.00 & 34.00 & 32.00 & 26.42 & 48.00 & 38.00 & 62.00 & 62.00 & 68.00 & 60.00 & 28.00 & 52.00 & 56.00 & 48.61 \\
DeepSeek-VL2 & 22.00 & 10.00 & 10.00 & 26.42 & 34.00 & 8.00 & 20.00 & 24.00 & 38.00 & 24.00 & 8.00 & 10.00 & 28.00 & 24.33 \\
Gemini-2.0 & \textbf{78.00} & 14.00 & 52.00 & \textbf{37.74} & \textbf{80.00} & \textbf{62.00} & \textbf{90.00} & {\ul 78.00} & {\ul 76.00} & \textbf{90.00} & 50.00 & \textbf{74.00} & {\ul 82.00} & \textbf{75.56} \\
GLM-4V-plus & 62.00 & 44.00 & 36.00 & 28.30 & 68.00 & 48.00 & 50.00 & 70.00 & 66.00 & 76.00 & 40.00 & 54.00 & 52.00 & 61.34 \\
GPT-4o & 54.00 & {\ul 58.00} & {\ul 54.00} & {\ul 33.96} & {\ul 74.00} & {\ul 56.00} & {\ul 80.00} & \textbf{98.00} & \textbf{92.00} & \textbf{90.00} & \textbf{68.00} & {\ul 70.00} & \textbf{92.00} & {\ul 71.18} \\
Intern-VL2.5-8B-MPO & 32.00 & 18.00 & 18.00 & 32.08 & 48.00 & 22.00 & 36.00 & 38.00 & 50.00 & 38.00 & 10.00 & 24.00 & 24.00 & 37.77 \\
Intern-VL2.5-78B-MPO & 58.00 & 52.00 & \textbf{58.00} & 28.30 & 60.00 & 46.00 & 74.00 & 70.00 & 74.00 & {\ul 84.00} & 50.00 & 58.00 & 72.00 & 66.03 \\
LLaVA-OneVision-7B & 4.00 & 0.00 & 0.00 & 5.66 & 10.00 & 8.00 & 12.00 & 4.00 & 76.00 & 60.00 & 2.00 & 0.00 & 4.00 & 11.78 \\
LLaVA-OneVision-72B & 20.00 & 14.00 & 2.00 & 28.30 & 28.00 & 14.00 & 30.00 & 14.00 & 62.00 & 44.00 & 14.00 & 22.00 & 38.00 & 29.61 \\
MiniCPM-V-2\_6 & 24.00 & 0.00 & 4.00 & 15.09 & 32.00 & 6.00 & 20.00 & 20.00 & 36.00 & 24.00 & 10.00 & 8.00 & 18.00 & 22.56 \\
Qwen2-VL-72B & {\ul 74.00} & \textbf{68.00} & 48.00 & 32.08 & 70.00 & 48.00 & 76.00 & 70.00 & 74.00 & {\ul 84.00} & {\ul 66.00} & 62.00 & 78.00 & 69.61 \\
Qwen2-VL-7B & 34.00 & 12.00 & 6.00 & 24.53 & 40.00 & 6.00 & 38.00 & 38.00 & 48.00 & 46.00 & 20.00 & 14.00 & 50.00 & 32.55 \\
Qwen2-VL-7B-FT & 64.00 & 36.00 & 38.00 & 20.75 & 62.00 & 18.00 & 56.00 & 70.00 & 50.00 & 68.00 & 48.00 & 40.00 & 0.00 & 54.08 \\ \bottomrule
\end{tabular}%
}
  \caption{\label{UML pass@1 UML flowchart}
    Pass@1 results for 13 LLMs on UML flowchart generation task in the Flow2Code benchmarks. 
    The results in bold are the optimal results, while the underlined results represent the suboptimal results.
    The results are represented in percentage (\%).
  }
\end{table*}

\begin{table*}[]
\resizebox{\textwidth}{!}{%
\begin{tabular}{@{}lcccccccccccccc@{}}
\toprule
\multicolumn{1}{l|}{\multirow{2}{*}{Model}} & \multicolumn{1}{c|}{ClassEval (100)} & \multicolumn{4}{c|}{HumanEval-X (164)} & \multicolumn{7}{c|}{MBXP (611)} & \multicolumn{2}{c}{McEval (50)} \\ \cmidrule(l){2-15} 
\multicolumn{1}{l|}{} & \multicolumn{1}{c|}{Python} & CPP & Java & JS & \multicolumn{1}{c|}{Python} & CPP & C\# & Java & JS & PHP & Python & \multicolumn{1}{c|}{Ruby} & C & C\# \\ \midrule
Claude-3.5-Sonnet & 2.00 & 62.80 & \textbf{71.34} & 43.29 & 73.17 & 51.72 & 66.12 & 48.12 & 80.85 & \textbf{100.00} & 80.69 & 81.51 & 58.00 & 66.00 \\
DeepSeek-VL2 & 1.00 & 0.00 & 4.27 & 19.51 & 36.59 & 1.80 & 0.82 & 1.80 & 36.82 & 60.23 & 36.99 & 40.92 & 14.00 & 12.00 \\
Gemini-2.0 & \textbf{44.00} & \textbf{78.05} & {\ul 70.73} & \textbf{86.59} & \textbf{85.37} & 50.74 & \textbf{79.54} & \textbf{80.52} & \textbf{90.67} & 77.09 & \textbf{92.80} & \textbf{87.23} & \textbf{76.00} & \textbf{88.00} \\
GLM-4V-plus & 1.00 & 20.12 & 7.93 & 29.88 & 73.17 & 28.97 & 21.11 & 4.91 & 76.27 & 71.69 & 66.61 & 79.54 & 50.00 & 40.00 \\
GPT-4o & 9.00 & 56.71 & 57.93 & 35.98 & \textbf{85.37} & {\ul 63.83} & 52.21 & 36.66 & 81.01 & {\ul 98.20} & 87.56 & 82.16 & {\ul 66.00} & {\ul 80.00} \\
Intern-VL2.5-8B-MPO & 7.00 & 0.00 & 3.66 & 43.29 & 54.88 & 1.80 & 11.95 & 5.07 & 60.56 & 93.94 & 61.70 & 60.07 & 26.00 & 8.00 \\
Intern-VL2.5-78B-MPO & {\ul 26.00} & 51.22 & 7.93 & {\ul 79.27} & 82.93 & 41.90 & 58.27 & 7.86 & 79.38 & 82.65 & 82.98 & {\ul 82.49} & 58.00 & 56.00 \\
LLaVA-OneVision-7B & 0.20 & 0.00 & 2.44 & 7.93 & 27.44 & 0.00 & 0.00 & 0.98 & 36.66 & 25.53 & 23.24 & 31.26 & 4.00 & 8.00 \\
LLaVA-OneVision-72B & 9.00 & 4.88 & 12.20 & 41.46 & 58.54 & 6.71 & 17.84 & 4.75 & 59.74 & \textbf{100.00} & 53.68 & 60.07 & 20.00 & 34.00 \\
MiniCPM-V-2\_6 & 0.00 & 0.00 & 0.61 & 6.71 & 37.80 & 0.00 & 0.82 & 0.49 & 38.30 & 35.68 & 37.64 & 39.44 & 12.00 & 8.00 \\
Qwen2-VL-72B & 19.00 & {\ul 64.63} & 23.78 & 73.17 & {\ul 84.15} & 40.75 & 67.43 & 9.66 & {\ul 83.80} & 98.04 & {\ul 84.62} & 81.83 & 62.00 & 74.00 \\
Qwen2-VL-7B & 2.00 & 0.00 & 22.56 & 39.02 & 59.15 & 1.47 & 1.47 & 12.27 & 62.19 & 65.47 & 59.57 & 28.81 & 26.00 & 42.00 \\
Qwen2-VL-7B-FT & 4.00 & 46.95 & 57.32 & 53.05 & 61.59 & \textbf{79.05} & {\ul 79.38} & {\ul 77.74} & 80.03 & 75.61 & 71.36 & 75.29 & 40.00 & 54.00 \\ \midrule
\multicolumn{1}{l|}{\multirow{2}{*}{Model}} & \multicolumn{14}{c}{McEval (50)} \\ \cmidrule(l){2-15} 
\multicolumn{1}{l|}{} & \multicolumn{1}{c|}{CPP} & \multicolumn{1}{c|}{Fortran} & \multicolumn{1}{c|}{HTML} & \multicolumn{1}{c|}{Java} & \multicolumn{1}{c|}{JS} & \multicolumn{1}{c|}{Pascal} & \multicolumn{1}{c|}{Perl} & \multicolumn{1}{c|}{PHP} & \multicolumn{1}{c|}{Python} & \multicolumn{1}{c|}{Ruby} & \multicolumn{1}{c|}{Shell} & \multicolumn{1}{c|}{Tcl} & \multicolumn{1}{c|}{VB} & Avg \\ \midrule
Claude-3.5-Sonnet & 52.00 & 26.00 & 12.00 & 30.19 & {\ul 72.00} & 52.00 & {\ul 72.00} & 70.00 & 80.00 & 80.00 & {\ul 50.00} & \textbf{78.00} & 80.00 & 60.73 \\
DeepSeek-VL2 & 22.00 & 2.00 & 0.00 & 18.87 & 34.00 & 0.00 & 26.00 & 28.00 & 32.00 & 34.00 & 8.00 & 4.00 & 18.00 & 18.69 \\
Gemini-2.0 & \textbf{72.00} & 30.00 & 20.00 & \textbf{15.09} & \textbf{78.00} & \textbf{68.00} & 68.00 & {\ul 84.00} & {\ul 88.00} & {\ul 86.00} & 44.00 & {\ul 74.00} & {\ul 82.00} & \textbf{70.98} \\
GLM-4V-plus & 50.00 & 24.00 & 16.00 & 45.28 & 56.00 & 38.00 & 46.00 & 70.00 & 52.00 & 76.00 & 22.00 & 32.00 & 66.00 & 44.17 \\
GPT-4o & 48.00 & {\ul 40.00} & {\ul 22.00} & \textbf{28.30} & 72.00 & {\ul 54.00} & \textbf{74.00} & \textbf{92.00} & \textbf{90.00} & \textbf{90.00} & 44.00 & 68.00 & \textbf{86.00} & {\ul 63.05} \\
Intern-VL2.5-8B-MPO & 36.00 & 12.00 & 8.00 & 26.42 & 42.00 & 10.00 & 26.00 & 28.00 & 28.00 & 36.00 & 14.00 & 10.00 & 30.00 & 27.77 \\
Intern-VL2.5-78B-MPO & 48.00 & \textbf{48.00} & 16.00 & 52.83 & 56.00 & 38.00 & 66.00 & 64.00 & 70.00 & 70.00 & 46.00 & 56.00 & 72.00 & 56.19 \\
LLaVA-OneVision-7B & 8.00 & 0.00 & 0.00 & 9.43 & 20.00 & 2.00 & 6.00 & 20.00 & 36.00 & 56.00 & 4.00 & 4.00 & 12.00 & 12.78 \\
LLaVA-OneVision-72B & 24.00 & 20.00 & 8.00 & {\ul 58.49} & 48.00 & 24.00 & 38.00 & 36.00 & 54.00 & 58.00 & 34.00 & 22.00 & 44.00 & 35.24 \\
MiniCPM-V-2\_6 & 12.00 & 0.00 & 0.00 & 13.00 & 22.00 & 2.00 & 10.00 & 36.00 & 28.00 & 18.00 & 18.00 & 16.00 & 6.00 & 16.43 \\
Qwen2-VL-72B & {\ul 60.00} & 32.00 & \textbf{26.00} & \textbf{66.04} & 66.00 & 46.00 & 60.00 & 78.00 & 66.00 & 80.00 & \textbf{52.00} & 56.00 & 58.00 & 59.81 \\
Qwen2-VL-7B & 30.00 & 2.00 & 10.00 & 22.64 & 48.00 & 4.00 & 30.00 & 48.00 & 44.00 & 42.00 & 24.00 & 12.00 & 30.00 & 29.12 \\
Qwen2-VL-7B-FT & 42.00 & 16.00 & 4.00 & 15.09 & 52.00 & 10.00 & 38.00 & 56.00 & 42.00 & 56.00 & 32.00 & 20.00 & 52.00 & 47.79 \\ \bottomrule
\end{tabular}%
}
  \caption{\label{UML pass@1 pseudocode flowchart}
    Pass@1 results for 13 LLMs on the pseudocode flowchart generation task in the Flow2Code benchmarks. 
    The results in bold are the optimal results, while the underlined results represent the suboptimal results.
    The results are represented in percentage (\%).
  }
\end{table*}

\begin{table*}[]
\resizebox{\textwidth}{!}{%
\begin{tabular}{@{}cccccccccccccc@{}}
\toprule
\multicolumn{1}{c|}{\multirow{2}{*}{Model}} & \multicolumn{1}{c|}{ClassEval (100)} & \multicolumn{4}{c|}{HumanEval-X (164)} & \multicolumn{7}{c|}{MBXP (611)} & \multirow{2}{*}{Avg} \\ \cmidrule(lr){2-13}
\multicolumn{1}{c|}{} & \multicolumn{1}{c|}{Python} & CPP & Java & JS & \multicolumn{1}{c|}{Python} & CPP & C\# & Java & JS & PHP & Python & \multicolumn{1}{c|}{Ruby} &  \\ \midrule
Claude-3.5-Sonnet & 33.60 & 45.24 & 56.10 & 49.27 & 59.02 & 58.64 & 63.24 & 60.03 & 59.07 & \textbf{100.00} & 68.56 & 59.20 & 59.33 \\
DeepSeek-VL2 & 3.40 & 3.72 & 32.26 & 54.70 & 65.61 & 35.30 & 29.30 & 27.99 & 71.05 & 99.12 & 69.84 & 72.98 & 47.15 \\
Gemini-2.0 & \textbf{73.90} & \textbf{90.67} & \textbf{93.54} & \textbf{90.85} & \textbf{91.65} & \textbf{92.39} & {\ul 89.80} & \textbf{92.08} & \textbf{94.53} & 96.15 & \textbf{95.22} & \textbf{96.63} & \textbf{91.77} \\
GLM-4V-plus & 20.70 & 61.46 & 65.55 & 80.55 & 76.77 & 78.33 & 72.68 & 77.94 & 84.14 & 98.81 & 81.85 & 86.84 & 73.90 \\
GPT-4o & {\ul 64.10} & 57.99 & {\ul 91.22} & {\ul 89.27} & {\ul 89.09} & 83.29 & 85.16 & 85.17 & 90.18 & \textbf{100.00} & {\ul 93.29} & {\ul 95.27} & {\ul 85.33} \\
Intern-VL2.5-8B-MPO & 21.90 & 24.94 & 30.49 & 65.37 & 67.68 & 46.25 & 28.43 & 31.11 & 69.38 & {\ul 99.98} & 74.12 & 79.26 & 53.24 \\
Intern-VL2.5-78B-MPO & 37.70 & 60.98 & 75.06 & 83.05 & 86.83 & 78.90 & 78.87 & 67.58 & 85.09 & 99.66 & 88.72 & 92.03 & 77.90 \\
LLaVA-OneVision-7B & 0.60 & 0.00 & 10.06 & 9.15 & 16.77 & 0.00 & 0.00 & 5.06 & 33.57 & 88.71 & 19.44 & 28.72 & 17.67 \\
LLaVA-OneVision-72B & 10.60 & 17.68 & 35.67 & 46.65 & 50.55 & 37.92 & 28.27 & 34.65 & 57.68 & \textbf{100.00} & 58.25 & 64.71 & 45.22 \\
MiniCPM-V-2\_6 & 3.00 & 1.10 & 22.50 & 35.18 & 44.57 & 12.93 & 19.39 & 9.95 & 58.92 & 96.73 & 52.70 & 70.61 & 35.88 \\
Qwen2-VL-72B & 50.90 & 62.74 & 77.56 & 79.02 & 81.71 & 81.46 & 76.50 & 75.58 & 86.25 & \textbf{100.00} & 86.46 & 90.33 & 79.04 \\
Qwen2-VL-7B & 20.40 & 4.39 & 59.94 & 67.44 & 77.20 & 20.90 & 9.17 & 33.88 & 75.22 & 88.90 & 71.42 & 51.82 & 49.11 \\
Qwen2-VL-7B-FT & 36.30 & {\ul 67.01} & 82.68 & 83.11 & 86.95 & {\ul 91.67} & \textbf{90.57} & {\ul 91.00} & {\ul 90.29} & 92.00 & 88.66 & 92.18 & 82.74 \\ \bottomrule
\end{tabular}%
}
  \caption{\label{flowchart pass@3}
    Pass@3 results for 13 LLMs on the code flowchart generation task in the Flow2Code benchmarks. 
    The results in bold are the optimal results, while the underlined results represent the suboptimal results.
    The results are represented in percentage (\%).
  }
\end{table*}

\begin{table*}[]
\resizebox{\textwidth}{!}{%
\begin{tabular}{@{}cccccccccccccc@{}}
\toprule
\multicolumn{1}{c|}{\multirow{2}{*}{Model}} & \multicolumn{1}{c|}{ClassEval (100)} & \multicolumn{4}{c|}{HumanEval-X (164)} & \multicolumn{7}{c|}{MBXP (611)} & \multirow{2}{*}{Avg} \\ \cmidrule(lr){2-13}
\multicolumn{1}{c|}{} & \multicolumn{1}{c|}{Python} & CPP & Java & JS & \multicolumn{1}{c|}{Python} & CPP & C\# & Java & JS & PHP & Python & \multicolumn{1}{c|}{Ruby} &  \\ \midrule
Claude-3.5-Sonnet & 30.70 & 36.59 & 51.71 & 28.66 & 61.59 & 60.98 & 66.22 & 63.96 & 59.87 & \textbf{100.00} & 67.56 & 61.42 & 57.44 \\
DeepSeek-VL2 & 8.50 & 7.38 & 28.78 & 61.77 & 61.04 & 43.65 & 29.31 & 26.94 & 68.92 & 98.59 & 65.55 & 73.26 & 47.88 \\
Gemini-2.0 & \textbf{67.40} & \textbf{78.60} & \textbf{83.72} & {\ul 89.57} & \textbf{88.84} & {\ul 90.07} & \textbf{91.33} & {\ul 86.99} & \textbf{92.50} & 91.90 & \textbf{93.76} & \textbf{95.02} & \textbf{88.15} \\
GLM-4V-plus & 18.30 & 57.56 & 63.54 & 82.99 & 78.23 & 79.77 & 69.87 & 76.74 & 83.26 & 99.15 & 79.38 & 85.73 & 72.95 \\
GPT-4o & {\ul 60.30} & 60.00 & {\ul 80.91} & \textbf{94.82} & {\ul 88.11} & 83.99 & 87.35 & 84.35 & {\ul 88.92} & {\ul 99.98} & {\ul 92.23} & {\ul 94.81} & {\ul 84.65} \\
Intern-VL2.5-8B-MPO & 18.00 & 23.11 & 34.21 & 73.96 & 63.54 & 51.03 & 37.25 & 42.18 & 70.38 & 99.74 & 72.19 & 74.75 & 55.05 \\
Intern-VL2.5-78B-MPO & 44.00 & 48.17 & 74.02 & 86.65 & 81.10 & 78.23 & 80.16 & 70.28 & 85.22 & 99.77 & 87.73 & 91.00 & 77.21 \\
LLaVA-OneVision-7B & 0.60 & 0.00 & 5.37 & 17.38 & 12.80 & 0.20 & 0.00 & 5.97 & 33.27 & 87.87 & 18.25 & 25.66 & 17.28 \\
LLaVA-OneVision-72B & 10.90 & 13.17 & 33.54 & 51.52 & 50.12 & 31.10 & 29.48 & 36.76 & 56.43 & \textbf{100.00} & 57.12 & 62.64 & 44.40 \\
MiniCPM-V-2\_6 & 5.30 & 3.05 & 19.88 & 53.66 & 43.35 & 23.40 & 29.85 & 4.12 & 56.55 & 97.94 & 55.34 & 69.95 & 38.69 \\
Qwen2-VL-72B & 44.10 & 58.05 & 76.71 & 78.90 & 78.29 & 81.26 & 76.07 & 77.58 & 85.24 & 99.93 & 84.39 & 91.00 & 77.63 \\
Qwen2-VL-7B & 20.60 & 10.06 & 51.10 & 66.16 & 67.87 & 50.98 & 13.85 & 21.19 & 73.26 & 91.62 & 73.58 & 57.87 & 50.32 \\
Qwen2-VL-7B-FT & 30.00 & {\ul 65.43} & 80.79 & 71.52 & 79.45 & \textbf{91.72} & {\ul 90.36} & \textbf{90.07} & 88.71 & 93.08 & 83.11 & 91.41 & 79.68 \\ \bottomrule
\end{tabular}%
}
  \caption{\label{UML pass@3}
    Pass@3 results for 13 LLMs on the UML flowchart generation task in the Flow2Code benchmarks. 
    The results in bold are the optimal results, while the underlined results represent the suboptimal results.
    The results are represented in percentage (\%).
  }
\end{table*}

\begin{table*}[]
\resizebox{\textwidth}{!}{%
\begin{tabular}{@{}cccccccccccccc@{}}
\toprule
\multicolumn{1}{c|}{\multirow{2}{*}{Model}} & \multicolumn{1}{c|}{ClassEval (100)} & \multicolumn{4}{c|}{HumanEval-X (164)} & \multicolumn{7}{c|}{MBXP (611)} & \multirow{2}{*}{Avg} \\ \cmidrule(lr){2-13}
\multicolumn{1}{c|}{} & \multicolumn{1}{c|}{Python} & CPP & Java & JS & \multicolumn{1}{c|}{Python} & CPP & C\# & Java & JS & PHP & Python & \multicolumn{1}{c|}{Ruby} &  \\ \midrule
Claude-3.5-Sonnet & 3.00 & 71.10 & {\ul 79.76} & 50.61 & 81.89 & 62.90 & 73.00 & 69.13 & 84.55 & \textbf{100.00} & 86.24 & 85.30 & 70.62 \\
DeepSeek-VL2 & 1.20 & 3.29 & 8.84 & 43.66 & 53.29 & 5.94 & 3.85 & 7.86 & 65.74 & 87.69 & 63.42 & 67.09 & 35.07 \\
Gemini-2.0 & \textbf{52.50} & \textbf{87.13} & \textbf{84.51} & \textbf{87.32} & {\ul 88.48} & 58.81 & {\ul 83.06} & \textbf{88.63} & \textbf{92.90} & 79.59 & \textbf{94.60} & \textbf{89.36} & \textbf{83.94} \\
GLM-4V-plus & 1.00 & 31.40 & 12.20 & 31.71 & 76.28 & 36.37 & 25.52 & 8.43 & 78.09 & 76.01 & 69.53 & 81.33 & 45.98 \\
GPT-4o & 18.00 & {\ul 79.02} & 79.21 & 50.55 & \textbf{91.10} & {\ul 80.77} & 77.66 & 69.44 & 87.86 & {\ul 99.93} & {\ul 92.05} & {\ul 88.59} & {\ul 76.19} \\
Intern-VL2.5-8B-MPO & 13.60 & 0.37 & 8.72 & 55.18 & 65.55 & 4.17 & 21.59 & 12.14 & 69.12 & 98.77 & 69.59 & 72.08 & 41.00 \\
Intern-VL2.5-78B-MPO & {\ul 33.00} & 67.80 & 16.34 & {\ul 83.54} & 83.23 & 50.54 & 68.41 & 17.04 & 83.65 & 93.06 & 86.45 & 86.38 & 64.70 \\
LLaVA-OneVision-7B & 0.60 & 0.00 & 7.74 & 22.26 & 41.46 & 0.00 & 0.00 & 3.34 & 52.57 & 50.00 & 38.67 & 48.90 & 22.13 \\
LLaVA-OneVision-72B & 9.30 & 10.49 & 31.34 & 53.84 & 69.51 & 10.74 & 35.34 & 14.27 & 67.91 & \textbf{100.00} & 65.86 & 74.06 & 45.22 \\
MiniCPM-V-2\_6 & 1.50 & 0.00 & 1.71 & 16.16 & 44.33 & 0.10 & 1.59 & 0.88 & 56.60 & 62.19 & 53.32 & 54.75 & 27.49 \\
Qwen2-VL-72B & 21.10 & 69.09 & 30.12 & 75.85 & 86.52 & 43.73 & 72.27 & 15.14 & 86.56 & 99.31 & 87.33 & 84.75 & 64.37 \\
Qwen2-VL-7B & 4.80 & 0.37 & 39.09 & 58.48 & 73.72 & 2.55 & 3.63 & 29.35 & 78.85 & 82.06 & 76.09 & 50.62 & 42.90 \\
Qwen2-VL-7B-FT & 9.00 & 61.59 & 73.90 & 64.09 & 79.63 & \textbf{87.74} & \textbf{89.75} & {\ul 87.04} & {\ul 88.99} & 85.40 & 88.35 & 86.84 & 75.22 \\ \bottomrule
\end{tabular}%
}
  \caption{\label{pseudocode pass@3}
    Pass@3 results for 13 LLMs on the pseudocode flowchart generation task in the Flow2Code benchmarks. 
    The results in bold are the optimal results, while the underlined results represent the suboptimal results.
    The results are represented in percentage (\%).
  }
\end{table*}

\begin{table*}[]
\resizebox{\textwidth}{!}{%
\begin{tabular}{@{}cccccccccccccc@{}}
\toprule
\multicolumn{1}{c|}{\multirow{2}{*}{Model}} & \multicolumn{1}{c|}{ClassEval (100)} & \multicolumn{4}{c|}{HumanEval-X (164)} & \multicolumn{7}{c|}{MBXP (611)} & \multirow{2}{*}{Avg} \\ \cmidrule(lr){2-13}
\multicolumn{1}{c|}{} & \multicolumn{1}{c|}{Python} & CPP & Java & JS & \multicolumn{1}{c|}{Python} & CPP & C\# & Java & JS & PHP & Python & \multicolumn{1}{c|}{Ruby} &  \\ \midrule
Claude-3.5-Sonnet & 37.00 & 50.61 & 58.54 & 53.05 & 61.59 & 63.50 & 66.78 & 63.67 & 62.68 & \textbf{100.00} & 71.85 & 63.18 & 62.70 \\
DeepSeek-VL2 & 4.00 & 4.88 & 42.07 & 60.98 & 71.34 & 47.14 & 39.61 & 38.46 & 78.07 & {\ul 99.84} & 76.43 & 80.20 & 53.60 \\
Gemini-2.0 & \textbf{75.00} & \textbf{92.07} & \textbf{93.90} & \textbf{91.46} & \textbf{92.07} & {\ul 93.13} & {\ul 91.16} & {\ul 92.96} & \textbf{95.25} & 98.04 & \textbf{95.58} & \textbf{97.22} & \textbf{92.48} \\
GLM-4V-plus & 22.00 & 64.02 & 66.46 & 81.10 & 77.44 & 78.89 & 73.65 & 78.56 & 84.62 & 99.02 & 82.65 & 87.40 & 74.73 \\
GPT-4o & {\ul 66.00} & 67.07 & {\ul 92.68} & {\ul 90.24} & {\ul 90.24} & 86.91 & 87.40 & 87.89 & 91.49 & \textbf{100.00} & {\ul 94.27} & {\ul 96.40} & {\ul 87.55} \\
Intern-VL2.5-8B-MPO & 26.00 & 29.27 & 35.37 & 67.68 & 69.51 & 51.55 & 33.88 & 35.84 & 72.67 & \textbf{100.00} & 75.94 & 81.67 & 56.62 \\
Intern-VL2.5-78B-MPO & 41.00 & 64.02 & 78.05 & 84.76 & 87.80 & 80.85 & 80.85 & 70.87 & 86.58 & 99.67 & 89.69 & 92.80 & 79.77 \\
LLaVA-OneVision-7B & 1.00 & 0.00 & 12.80 & 13.41 & 20.12 & 0.00 & 0.00 & 7.36 & 40.43 & 93.78 & 24.22 & 35.19 & 20.69 \\
LLaVA-OneVision-72B & 12.00 & 21.95 & 39.63 & 50.61 & 54.88 & 44.84 & 34.70 & 43.37 & 61.54 & \textbf{100.00} & 63.01 & 69.07 & 49.63 \\
MiniCPM-V-2\_6 & 3.00 & 1.83 & 26.83 & 41.46 & 48.17 & 17.35 & 25.20 & 14.89 & 63.83 & 98.20 & 57.12 & 76.10 & 39.64 \\
Qwen2-VL-72B & 55.00 & 65.24 & 79.27 & 80.49 & 82.93 & 83.14 & 78.40 & 78.23 & 87.23 & \textbf{100.00} & 87.73 & 91.00 & 80.72 \\
Qwen2-VL-7B & 26.00 & 7.32 & 68.90 & 71.95 & 81.71 & 29.79 & 13.09 & 46.15 & 80.69 & 94.27 & 76.60 & 61.05 & 55.20 \\
Qwen2-VL-7B-FT & 41.00 & {\ul 71.34} & 85.98 & 84.76 & 89.63 & \textbf{93.45} & \textbf{91.98} & \textbf{93.13} & {\ul 92.14} & 92.80 & 91.00 & 93.62 & 85.12 \\ \bottomrule
\end{tabular}%
}
  \caption{\label{flowchart pass@5}
    Pass@5 results for 13 LLMs on the code flowchart generation task in the Flow2Code benchmarks. 
    The results in bold are the optimal results, while the underlined results represent the suboptimal results.
    The results are represented in percentage (\%).
  }
\end{table*}

\begin{table*}[]
\resizebox{\textwidth}{!}{%
\begin{tabular}{@{}cccccccccccccc@{}}
\toprule
\multicolumn{1}{c|}{\multirow{2}{*}{Model}} & \multicolumn{1}{c|}{ClassEval (100)} & \multicolumn{4}{c|}{HumanEval-X (164)} & \multicolumn{7}{c|}{MBXP (611)} & \multirow{2}{*}{Avg} \\ \cmidrule(lr){2-13}
\multicolumn{1}{c|}{} & \multicolumn{1}{c|}{Python} & CPP & Java & JS & \multicolumn{1}{c|}{Python} & CPP & C\# & Java & JS & PHP & Python & \multicolumn{1}{c|}{Ruby} &  \\ \midrule
Claude-3.5-Sonnet & 35.00 & 40.24 & 55.49 & 31.10 & 63.41 & 64.98 & 70.38 & 67.76 & 63.99 & \textbf{100.00} & 70.87 & 64.98 & 60.68 \\
DeepSeek-VL2 & 11.00 & 10.37 & 36.59 & 66.46 & 66.46 & 53.19 & 39.12 & 36.50 & 77.25 & 99.51 & 71.52 & 80.03 & 54.03 \\
Gemini-2.0 & \textbf{68.00} & \textbf{81.10} & \textbf{84.76} & {\ul 90.24} & \textbf{90.24} & {\ul 90.83} & \textbf{92.14} & {\ul 88.87} & \textbf{93.13} & 94.44 & \textbf{94.27} & {\ul 95.58} & \textbf{89.10} \\
GLM-4V-plus & 19.00 & 60.37 & 64.63 & 83.54 & 79.27 & 80.52 & 71.36 & 77.74 & 84.12 & 99.35 & 80.20 & 85.92 & 73.89 \\
GPT-4o & {\ul 64.00} & 67.68 & {\ul 83.54} & \textbf{96.34} & {\ul 89.63} & 87.23 & {\ul 89.53} & 86.74 & {\ul 90.83} & \textbf{100.00} & {\ul 93.29} & \textbf{96.07} & {\ul 87.07} \\
Intern-VL2.5-8B-MPO & 21.00 & 27.44 & 37.80 & 76.83 & 67.68 & 55.97 & 43.21 & 48.12 & 74.80 & {\ul 99.84} & 75.45 & 77.09 & 58.78 \\
Intern-VL2.5-78B-MPO & 47.00 & 50.00 & 76.22 & 87.80 & 82.32 & 80.69 & 82.49 & 73.16 & 87.07 & {\ul 99.84} & 89.03 & 91.98 & 78.98 \\
LLaVA-OneVision-7B & 1.00 & 0.00 & 7.93 & 22.56 & 15.24 & 0.33 & 0.00 & 8.67 & 40.10 & 93.78 & 21.93 & 31.75 & 20.27 \\
LLaVA-OneVision-72B & 12.00 & 16.46 & 37.80 & 56.71 & 55.49 & 36.01 & 36.01 & 45.01 & 61.05 & \textbf{100.00} & 61.21 & 67.92 & 48.81 \\
MiniCPM-V-2\_6 & 7.00 & 3.05 & 23.78 & 58.54 & 48.17 & 28.31 & 36.17 & 6.22 & 62.19 & 99.02 & 60.07 & 75.29 & 42.40 \\
Qwen2-VL-72B & 47.00 & 61.59 & 78.66 & 79.88 & 79.27 & 82.65 & 77.74 & 79.38 & 86.42 & \textbf{100.00} & 85.60 & 91.65 & 79.15 \\
Qwen2-VL-7B & 25.00 & 13.41 & 60.98 & 70.12 & 71.95 & 59.57 & 19.15 & 30.44 & 77.58 & 95.91 & 79.05 & 66.45 & 56.09 \\
Qwen2-VL-7B-FT & 32.00 & {\ul 69.51} & \textbf{84.76} & 74.39 & 84.76 & \textbf{93.62} & \textbf{92.14} & \textbf{92.14} & 90.67 & 95.09 & 86.58 & 93.13 & 82.45 \\ \bottomrule
\end{tabular}%
}
  \caption{\label{UML pass@5}
    Pass@5 results for 13 LLMs on the UML flowchart generation task in the Flow2Code benchmarks. 
    The results in bold are the optimal results, while the underlined results represent the suboptimal results.
    The results are represented in percentage (\%).
  }
\end{table*}

\begin{table*}[]
\resizebox{\textwidth}{!}{%
\begin{tabular}{@{}cccccccccccccc@{}}
\toprule
\multicolumn{1}{c|}{\multirow{2}{*}{Model}} & \multicolumn{1}{c|}{ClassEval (100)} & \multicolumn{4}{c|}{HumanEval-X (164)} & \multicolumn{7}{c|}{MBXP (611)} & \multirow{2}{*}{Avg} \\ \cmidrule(lr){2-13}
\multicolumn{1}{c|}{} & \multicolumn{1}{c|}{Python} & CPP & Java & JS & \multicolumn{1}{c|}{Python} & CPP & C\# & Java & JS & PHP & Python & \multicolumn{1}{c|}{Ruby} &  \\ \midrule
Claude-3.5-Sonnet & 3.00 & 73.17 & 81.71 & 52.44 & 84.15 & 66.78 & 74.30 & 75.12 & 85.92 & \textbf{100.00} & 88.22 & 86.58 & 72.62 \\
DeepSeek-VL2 & 2.00 & 5.49 & 13.41 & 53.05 & 59.15 & 9.17 & 5.89 & 12.27 & 73.98 & 95.25 & 70.38 & 74.30 & 39.86 \\
Gemini-2.0 & \textbf{55.00} & \textbf{89.63} & \textbf{87.20} & \textbf{88.41} & {\ul 89.02} & 62.68 & {\ul 83.80} & \textbf{90.67} & \textbf{93.45} & 80.52 & \textbf{94.93} & \textbf{90.02} & \textbf{85.40} \\
GLM-4V-plus & 1.00 & 33.54 & 15.24 & 32.32 & 76.83 & 39.28 & 27.17 & 9.66 & 78.56 & 77.58 & 70.21 & 81.67 & 47.12 \\
GPT-4o & 21.00 & {\ul 82.93} & {\ul 84.76} & 55.49 & \textbf{92.07} & {\ul 84.94} & 80.36 & 78.56 & 89.20 & \textbf{100.00} & {\ul 93.29} & 89.53 & {\ul 79.34} \\
Intern-VL2.5-8B-MPO & 16.00 & 0.61 & 10.37 & 60.37 & 69.51 & 5.40 & 25.53 & 17.18 & 71.36 & 99.18 & 72.50 & 75.12 & 43.66 \\
Intern-VL2.5-78B-MPO & {\ul 36.00} & 70.73 & 20.12 & {\ul 85.98} & 84.15 & 53.68 & 70.70 & 23.57 & 84.94 & 95.74 & 87.40 & 87.56 & 67.07 \\
LLaVA-OneVision-7B & 1.00 & 0.00 & 11.59 & 30.49 & 48.17 & 0.00 & 0.00 & 4.91 & 59.41 & 61.05 & 44.03 & 57.12 & 26.48 \\
LLaVA-OneVision-72B & 12.00 & 14.63 & 41.46 & 57.93 & 75.00 & 13.42 & 42.39 & 21.11 & 70.87 & \textbf{100.00} & 70.38 & 78.23 & 49.79 \\
MiniCPM-V-2\_6 & 2.00 & 0.00 & 2.44 & 21.95 & 49.39 & 0.16 & 2.29 & 1.47 & 63.67 & 72.18 & 59.41 & 60.88 & 30.31 \\
Qwen2-VL-72B & 22.00 & 70.73 & 32.93 & 76.83 & 87.20 & 44.68 & 73.98 & 17.68 & 87.23 & {\ul 99.51} & 88.05 & 85.60 & 65.58 \\
Qwen2-VL-7B & 6.00 & 0.61 & 47.56 & 65.24 & 78.05 & 3.60 & 5.89 & 40.75 & 82.65 & 87.89 & 80.03 & 60.39 & 47.50 \\
Qwen2-VL-7B-FT & 11.00 & 67.68 & 78.05 & 69.51 & 85.37 & \textbf{89.69} & \textbf{91.98} & {\ul 89.85} & {\ul 91.82} & 87.23 & 91.82 & {\ul 89.85} & 78.70 \\ \bottomrule
\end{tabular}%
}
  \caption{\label{pseudocode pass@5}
    Pass@5 results for 13 LLMs on the pseudocode flowchart generation task in the Flow2Code benchmarks. 
    The results in bold are the optimal results, while the underlined results represent the suboptimal results.
    The results are represented in percentage (\%).
  }
\end{table*}

\end{CJK*}
\end{document}